\documentclass[nofootinbib]{revtex4-2}

\usepackage{geometry}
\usepackage{amsmath,amssymb,amsfonts,dcolumn,color,graphicx}
\usepackage{latexsym,placeins,epsfig,tikz,amsbsy,bm,enumitem,mathrsfs}
\usetikzlibrary{arrows,shapes}
\usepackage{subfigure,rotating}
\usepackage[title]{appendix}
\usepackage[pagebackref=false,hidelinks]{hyperref}

\allowdisplaybreaks

\newcommand{\cH}{\mathcal H}

\newcommand{\ue}{\mathrm{UE}}
\newcommand{\ud}{\mathrm{UD}}
\newcommand{\lgauge}{\mathrm{L}}
\newcommand{\sfgauge}{\mathrm{SF}}
\newcommand{\cgauge}{\mathrm{C}}

\newcommand{\Order}{\mathcal O}
\newcommand{\factLW}{\mathcal F_{\mathrm{LW}}}
\newcommand{\factSW}{\mathcal F_{\mathrm{SW}}}

\begin{document}

\title{Gauge and scale dependence of the quadratic backreaction source during inflation: Uniform-expansion gauge}

\author{Inyong Cho}
\email{iycho@seoultech.ac.kr}
\affiliation{School of Natural Sciences,
College of Liberal Arts,
Seoul National University of Science and Technology,
Seoul 01811, Republic of Korea}

\begin{abstract}
Cosmological backreaction is most directly expressed through the local expansion
of the spacetime foliation, so uniform-expansion (UE) slicing provides a natural
geometrical setting in which to examine the quadratic effective energy-momentum
tensor (2EMT) of scalar perturbations during slow-roll inflation.
We impose the complete scalar gauge conditions $\kappa=0$ and $E=0$.
The uniform-expansion condition gives a \textit{differential} constraint
for the shift $Q$ in configuration space and satisfies
$Q_{\mathrm{UE}}(k)=-3(\Psi'+\mathcal H\Psi)/(k^2+3\epsilon\mathcal H^2)$
for each Fourier mode.
In the strict infrared, the UE source vanishes at zero gradient,
while its first nonzero terms are
$\widehat\tau_{00}^{\mathrm{UE}}/\mathcal F_{\mathrm{LW}}
\simeq8\sigma_1^2/\epsilon$ and
$\widehat\tau_{ii}^{\mathrm{UE}}/\mathcal F_{\mathrm{LW}}
\simeq-8\sigma_1^2/(3\epsilon)$ at leading slow-roll order.
Thus $w_{\rm raw}^{\mathrm{UE}}\simeq-1/3$, but the first finite-gradient
coefficients differ from the comoving and uniform-density gauges.
Sufficiently deep in the gradient-dominated intermediate infrared,
the leading UE behavior approaches the spatially-flat result,
whereas the crossover region between the strict and intermediate
infrared regimes requires the full expressions for the 2EMT.
In the ultraviolet, we have
$\widehat\tau_{00}^{\mathrm{UE}}\simeq2\mathcal F_{\mathrm{SW}}$ and
$\widehat\tau_{ii}^{\mathrm{UE}}\simeq(2/3)\mathcal F_{\mathrm{SW}}$,
yielding $w_{\mathrm{raw}}\simeq1/3$.
Separately, we demonstrate the gauge dependence of the 2EMT
under general scalar gauge transformations and discuss the complete fixing
of the UE gauge.
\end{abstract}

\maketitle

\section{Introduction}
\label{sec:intro}

Cosmological backreaction is fundamentally a question about expansion:
how inhomogeneous perturbations alter the local expansion
and how these effects modify the effective evolution
relative to a homogeneous Friedmann background.
The expansion scalar $\Theta$ of the hypersurface-normal congruence
is the negative trace of the extrinsic curvature,
$\Theta=-\mathcal K$,
and therefore controls the local volume expansion
of the chosen foliation.
Among the complete scalar gauges that we have studied so far,
the uniform-expansion gauge is the one most directly adapted
to this geometrical aspect of the backreaction problem.
With the common spatial gauge choice $E=0$,
the temporal condition of the uniform-expansion gauge fixes
the perturbation of the extrinsic-curvature trace itself,
$\kappa=0$,
rather than a matter variable
($\delta\phi=0$ in the comoving gauge
or $\delta\rho=0$ in the uniform-density gauge),
the scalar shear ($\beta=0$ in the longitudinal gauge),
or the intrinsic spatial curvature
($\psi=0$ in the spatially-flat gauge).
Uniform-expansion slicing is closely related
to constant-mean-curvature slicing,
and has a natural role in nonlinear
and long-wavelength descriptions of cosmological inhomogeneity
\cite{Geshnizjani:2002wp,Brandenberger:2002sk,Geshnizjani:2003cn,Harada:2015yda}.
This direct geometrical connection does not by itself make
a gauge-fixed effective source an observable,
but it makes the uniform-expansion sector especially informative
when the object under discussion is cosmological backreaction.

At quadratic order the influence of linear perturbations
can be collected into a quadratic effective energy-momentum tensor (2EMT),
obtained by rearranging the second-order Einstein equation
so that genuine second-order fields appear linearly
while products of first-order perturbations act as an effective source
\cite{Mukhanov:1996ak,Abramo:1997hu}.
The interpretation of such quadratic sources
and of cosmological backreaction has long involved gauge subtleties
\cite{Unruh:1998ic,Ishibashi:2005sj,Green:2010qy}.
In our previous calculations the scalar 2EMT
was evaluated explicitly in several complete gauges,
revealing structured rather than arbitrary dependence
on slicing and threading
\cite{Cho:2020zbh,Cho:2022maa,Cho:2024vpv,Cho:2026gjy}.
For comparison, we use the results for the longitudinal,
spatially-flat, comoving, and uniform-density gauges obtained in
Ref.~\cite{Cho:2026gjy}, where the relevant background quantities
were expressed exactly in terms of the slow-roll parameters before
performing the slow-roll expansion. The UE results are evaluated
using the same procedure.

A second principal aim of the present work
is to address this gauge question directly
at the level of the general quadratic source
before any gauge condition is imposed.
The early 2EMT formulations were constructed
to describe backreaction in a gauge-invariant manner
\cite{Mukhanov:1996ak,Abramo:1997hu},
while subsequent analyses emphasized the gauge and observer dependence
of second-order effective sources and their physical interpretation
\cite{Unruh:1998ic,Ishibashi:2005sj,Green:2010qy}.
Previous calculations, including our own,
primarily diagnosed the issue by comparing results
obtained after different gauge choices.
Here, no temporal or spatial gauge condition is imposed initially.
We apply an arbitrary infinitesimal scalar gauge transformation
$\xi^\mu=(\xi^0,\partial^i\zeta)$ directly
to the general quadratic source in configuration space
and evaluate its component variation.
The $00$ and diagonal spatial components retain
nonvanishing even-in-$k$ contributions for general gauge functions,
and therefore the general scalar 2EMT before gauge fixing
is not gauge invariant.
To our knowledge, this is the first explicit component-level evaluation
of the general scalar-gauge variation of the quadratic source
in this 2EMT framework.
The calculation also separates two logically distinct questions:
gauge invariance of the general quadratic source before gauge fixing,
and uniqueness of the representation selected
by a complete gauge fixing.

The most recent member of that sequence
was the uniform-density analysis \cite{Cho:2026gjy}.
There the temporal slicing is defined
by a slowly evolving matter clock, $\rho$,
and the ultraviolet 2EMT contains a $1/\epsilon$
($\epsilon$: slow-roll parameter) enhancement
associated with the conditioning of that clock,
together with an additional $1/\sigma_2^2$
($\sigma_2\equiv\cH/k$: short-wavelength parameter)
gradient factor from the Laplacian contribution
to the density constraint.
The uniform-expansion gauge considered here is qualitatively different.
It is a geometrical-expansion clock rather than a matter clock,
and the solution of its constraint introduces
the characteristic combination
\begin{equation}
 k^2+3\epsilon\cH^2.
\label{eq:charfactor}
\end{equation}
Thus the new problem is controlled
by a competition between the perturbation-gradient scale $k^2$
and the slow-roll geometrical scale $\epsilon\cH^2$.

The UE results obtained in this work exhibit substantially
different infrared behavior from those found in the four previously
studied gauges.
At exactly zero gradient both UE source components vanish.
At small nonzero $k$, with
$\sigma_1\equiv k/\cH$ denoting the long-wavelength parameter,
their first nonzero terms are
\begin{equation}
 \frac{\widehat\tau_{00}^{\ue}}{\factLW}
 \simeq\frac{8}{\epsilon}\sigma_1^2,
 \qquad
 \frac{\widehat\tau_{ii}^{\ue}}{\factLW}
 \simeq-\frac{8}{3\epsilon}\sigma_1^2,
\end{equation}
at leading slow-roll order.
Here we define the diagnostic ratio
$w_{\mathrm{raw}}^{\ue}\equiv
\widehat\tau_{ii}^{\ue}/\widehat\tau_{00}^{\ue}$,
which approaches $-1/3$ in the strict infrared limit at leading
slow-roll order.
The first finite-gradient coefficients differ
from the comoving and uniform-density sectors.

On the gradient-dominated side,
sufficiently far from the transition
that inverse-$\sigma_1$ corrections are subleading,
the leading UE behavior approaches the spatially-flat sector.
The same denominator makes the asymptotic analysis genuinely nonuniform:
factors of $(k^2+3\epsilon\cH^2)^{-n}$ generate inequivalent series
in the strict infrared hierarchy $\sigma_1^2\ll\epsilon$
and in the slow-roll-first hierarchy
$\epsilon,|\delta|\ll\sigma_1^2\ll1$ ($\delta$: slow-roll parameter).
Inverse powers of $\sigma_1$ at higher slow-roll orders
are therefore not strict-infrared divergences;
they signal nonuniformity as the transition is approached.
The scale $\sigma_1^2\sim3\epsilon$
marks the crossover of the UE constraint denominator,
but it does not by itself define
a sharp crossover of the full 2EMT.
Near this region the full rational expressions,
rather than a finite slow-roll-first truncation,
must be used.

A second distinctive feature appears before the Fourier reduction.
The condition $\kappa=0$ is a \textit{differential}
constraint for the shift variable $Q\equiv\beta+E'$ in configuration space,
and the gauge-fixed 2EMT retains $Q$, $Q'$, and $Q''$
in an unreduced derivative representation.
In Fourier space the same constraint becomes an \textit{algebraic}
relation for each Fourier mode,
allowing $Q_{\bm k}$ to be eliminated
in favor of the Bardeen potential $\Psi_{\bm k}$.
This reduction is a convenient representation
of the UE gauge-fixed result,
but it is not by itself a gauge-invariance criterion.
Once the gauge is completely fixed,
equivalent variables may be used to rewrite
the same uniquely specified gauge-fixed quantity.

In the strict ultraviolet limit, the UE source has
the same leading coefficients as the longitudinal and
spatially-flat sectors,
$\widehat\tau_{00}^{\ue}\simeq2\factSW$
and $\widehat\tau_{ii}^{\ue}\simeq(2/3)\factSW$,
yielding $w_{\mathrm{raw}}^{\ue}\to1/3$.

The paper is organized as follows.
In Sec.~\ref{sec:setup} we review the scalar perturbations,
gauge-invariant variables, linear solutions,
wavelength hierarchies, and the definition of the 2EMT,
using the same conventions and presentation
as the uniform-density companion paper.
In Sec.~\ref{sec:UE} we first review
the meaning of a scalar gauge transformation
and the criterion for gauge invariance,
and then apply the general transformation directly
to the quadratic source before gauge fixing.
We next relate the local expansion
to the trace of the extrinsic curvature,
impose the uniform-expansion gauge,
derive the Fourier-space shift,
and discuss complete gauge fixing
and the uniqueness of the resulting gauge-fixed representation.
Sections~\ref{sec:UELW} and \ref{sec:UESW}
present the full long- and short-wavelength 2EMTs
and analyze the \textit{strict} and \textit{intermediate} regimes.
Section~\ref{sec:comparison} compares all five gauges
and develops the scale-transition interpretation,
including the e-fold evolution of a fixed mode.
Section~\ref{sec:discussion} discusses the interpretation
of the gauge-fixed result
and its distinction from a fully specified relational observable.
We conclude in Sec.~\ref{sec:conclusion}.
Appendix~\ref{app:rawGT} gives the explicit general scalar-gauge variation
of the quadratic source before gauge fixing,
while Appendix~\ref{app:unreduced} records the unreduced
uniform-expansion 2EMT before eliminating $Q$.
The lengthy general 2EMT
and the full expressions for the other four gauges
are not repeated here;
they are available in Refs.~\cite{Cho:2022maa,Cho:2026gjy}.
\section{Scalar perturbations and the quadratic effective source}
\label{sec:setup}

The perturbative setup, notation, linear equations, Fourier conventions, and definition of the quadratic source in
this section are the same as those given in Sec.~II of the uniform-density companion paper
Ref.~\cite{Cho:2026gjy}.  We reproduce them here for convenience and to keep the present analysis self-contained.

\subsection{Metric, matter field, and gauge transformations}

We consider scalar perturbations of a spatially flat Friedmann background,
\begin{equation}
 ds^2=a^2(\eta)\left\{-(1+2\alpha)d\eta^2-2\beta_{,i}d\eta\,dx^i
 +\left[(1-2\psi)\delta_{ij}+2E_{,ij}\right]dx^i dx^j\right\},
\label{eq:metric}
\end{equation}
and a canonical scalar field
\begin{equation}
 \phi(\eta,\bm x)=\phi_0(\eta)+\delta\phi(\eta,\bm x).
\end{equation}
The energy-momentum tensor is
\begin{equation}
 T_{\mu\nu}=\phi_{,\mu}\phi_{,\nu}-g_{\mu\nu}
 \left[\frac12g^{\rho\sigma}\phi_{,\rho}\phi_{,\sigma}+V(\phi)\right].
\end{equation}
The background equations are
\begin{align}
 \cH^2=\frac{8\pi G}{3}\left(\frac12\phi_0'^2+a^2V\right),
\end{align}
\begin{align}
 2\cH'+\cH^2=8\pi G\left(-\frac12\phi_0'^2+a^2V\right),
\end{align}
\begin{align}
 \phi_0''+2\cH\phi_0'+a^2V_\phi=0,
\end{align}
where $\cH=a'/a$ and a prime denotes $d/d\eta$.

Under the infinitesimal transformation $x^\mu\to x^\mu+\xi^\mu$
generated by $\xi^\mu=(\xi^0,\partial^i\zeta)$, the perturbation variables transform as
\begin{align}
 \widetilde\alpha&=\alpha-\xi^{0\prime}-\cH\xi^0,\\
 \widetilde\beta&=\beta-\xi^0+\zeta',\\
 \widetilde\psi&=\psi+\cH\xi^0,\\
 \widetilde E&=E-\zeta,\\
 \widetilde{\delta\phi}&=\delta\phi-\phi_0'\xi^0.
\end{align}
Introducing
\begin{equation}
 Q\equiv\beta+E',\qquad
 \widetilde Q=Q-\xi^0,
\end{equation}
we define the first-order gauge-invariant combinations in the standard scalar-perturbation framework
\cite{Mukhanov:1990me}
\begin{align}
 \Phi&=\alpha-Q'-\cH Q,\\
 \Psi&=\psi+\cH Q,\\
 \delta\phi_{\rm gi}&=\delta\phi-\phi_0'Q.
\end{align}
Here, $\Phi$ and $\Psi$ are the so called Bardeen potentials.
For a canonical scalar field without anisotropic
stress, $\Phi=\Psi$.  The first-order equations give
\begin{align}
 \delta\phi_{\rm gi}&=\frac{\Psi'+\cH\Psi}{4\pi G\phi_0'},\label{eq:dphigi}\\
 \delta\phi_{\rm gi}'&=\frac{\Delta\Psi-K\Psi'-L\Psi}{4\pi G\phi_0'},
\end{align}
where
\begin{equation}
 K=3\cH+a^2\frac{V_\phi}{\phi_0'},\qquad
 L=\cH'+2\cH^2+a^2\cH\frac{V_\phi}{\phi_0'}.
\end{equation}
The Bardeen potential obeys
\begin{equation}
 \Psi''-\Delta\Psi+2K\Psi'+2L\Psi=0.
\label{eq:PsiEOM}
\end{equation}

\subsection{Slow-roll and wavelength series expansions}

We use the slow-roll parameters
\begin{equation}
 \epsilon\equiv-\frac{\dot H}{H^2}
 =4\pi G\frac{\dot\phi_0^2}{H^2},
 \qquad
 \delta\equiv-\frac{\ddot\phi_0}{H\dot\phi_0}
 =\epsilon-\frac{\dot\epsilon}{2H\epsilon}.
\label{eq:slowroll}
\end{equation}
where $H=\dot a/a$
and an overdot denotes the cosmic-time derivative.
We introduce an additional dimensionless parameter
characterizing the third cosmic-time derivative of the background inflaton:
\begin{equation}
 \xi_3\equiv\frac{\dddot\phi_0}{H^2\dot\phi_0}.
\label{eq:xi3}
\end{equation}
This parameter appears in the exact evolution equation
for $\delta$, namely
$\dot\delta=H[\delta(\epsilon+\delta)-\xi_3]$,
as well as in the expression for $V_{\phi\phi}$.
In the slow-roll hierarchy, $\xi_3$ is of second order,
together with $\epsilon^2$, $\epsilon\delta$, and $\delta^2$.
The exact background relations used in our calculation are
\begin{equation}
\begin{aligned}
 \epsilon'&=2\cH\epsilon(\epsilon-\delta),
 &\delta'&=\cH[\delta(\epsilon+\delta)-\xi_3],\\
 \cH'&=(1-\epsilon)\cH^2,
 &\phi_0'&=-\sqrt{2\epsilon}\,M_p\cH,\\
 a^2V&=(3-\epsilon)M_p^2\cH^2,
 &a^2V_\phi&=\sqrt{2\epsilon}(3-\delta)M_p\cH^2,\\
 &&a^2V_{\phi\phi}&=\cH^2[3(\epsilon+\delta)-\xi_3].
\end{aligned}
\label{eq:bgconsistent}
\end{equation}
Here $M_p^{-2}=8\pi G$. Equivalently,
\begin{equation*}
 \frac{a^2 V_{\phi\phi}}{\cH^2}
 =(3-\delta)(\epsilon+\delta)+\frac{\delta'}{\cH}.
\end{equation*}
Although $\xi_3$ is second order in slow roll,
we retain it in the exact background relations until the final
slow-roll expansion. This is particularly important when comparing
with the comoving and uniform-density gauges, where inverse powers
of $\epsilon$ can promote nominally higher-order contributions
into lower-order terms of the 2EMT.
The conformal-time relation for $\cH$ can also be written as
\begin{equation}
 \cH'-\cH^2=-\epsilon\cH^2.
\label{eq:Hprime}
\end{equation}
The complete gauge-fixed 2EMT is assembled
using Eq.~\eqref{eq:bgconsistent}
before the slow-roll truncation is imposed.

We decompose the perturbation into Fourier modes using the convention
\begin{align}
 \Psi(\eta,\bm x)
 &=\int d^3k\,\Psi_{\bm k}(\eta)e^{i\bm k\cdot\bm x},
 \label{eq:FourierExpansion}\\
 \Psi_{\bm k}(\eta)
 &=\frac{1}{(2\pi)^3}\int d^3x\,\Psi(\eta,\bm x)e^{-i\bm k\cdot\bm x}.
 \label{eq:FourierTransform}
\end{align}
The first relation is the Fourier-mode expansion,
whereas the second transforms the configuration-space field
to momentum space.
For each Fourier mode,
the long- and short-wavelength solutions are written as
\cite{Cho:2022maa,Cho:2024vpv}
\begin{equation}
 \Psi_{\bm k}(\eta)\simeq
 \begin{cases}
 A_1\epsilon,
 & \bigl(k\ll\cH:\ \text{long-wavelength limit}\bigr),\\[1mm]
 \displaystyle 4\pi G\frac{\phi_0'}{a}
 \left[c_1\sin(k\eta)+c_2\cos(k\eta)\right],
 & \bigl(k\gg\cH:\ \text{short-wavelength limit}\bigr).
 \end{cases}
\label{eq:PsiSolutions}
\end{equation}
We use these as the leading asymptotic mode inputs.
Explicit $k^2$, $k^4$, and higher-gradient coefficients
are evaluated within this mode approximation.
For short-wavelength products we retain the slowly
varying envelope when differentiating, and only then take the
fast-phase average,
\begin{equation}
 \overline{\sin^2(k\eta)}
 =\overline{\cos^2(k\eta)}=\frac12,
 \qquad
 \overline{\sin(k\eta)\cos(k\eta)}=0.
\label{eq:phaseaverage}
\end{equation}

We introduce independent wavelength parameters
in the long- and short-wavelength domains respectively,
\begin{equation}
 \sigma_1\equiv\frac{k}{\cH}\ll1,
 \qquad
 \sigma_2\equiv\frac{\cH}{k}\ll1.
\label{eq:sigmas}
\end{equation}
The slow-roll and gradient limits must be tracked separately.
Throughout this paper we distinguish four asymptotic regimes.
The \emph{strict IR} and \emph{strict UV} limits
are obtained by performing the wavelength expansion first
in $\sigma_1$ or $\sigma_2$, respectively,
at fixed admissible nonzero slow-roll parameters.
The formal limit $\sigma_i\to0$ is taken afterward.
The \emph{intermediate IR} and \emph{intermediate UV} regimes are defined by
\begin{equation}
 \epsilon,|\delta|\ll\sigma_i^2\ll1,
\label{eq:intermediateHierarchy}
\end{equation}
and are organized by performing the slow-roll expansion first
and then according to the wavelength hierarchy.
After extracting the displayed overall factor,
we retain zeroth- and first-order slow-roll terms
in each wavelength coefficient.
We denote the coefficientwise slow-roll remainder
at fixed $\sigma_i$ by $\Order_{\rm SR}(2;\sigma_i)$.

For the four gauges analyzed previously,
the corresponding gauge-fixed Fourier-space 2EMT components
contain only a finite set of polynomial
or Laurent-polynomial terms in the wavelength parameters
after the asymptotic mode solutions are substituted
\cite{Cho:2026gjy}.
Thus, performing a wavelength or slow-roll expansion
merely reorders the terms already present in the 2EMT,
and does not generate an additional infinite sequence.

The uniform-expansion gauge is qualitatively different
because its exact expressions contain factors
of $(k^2+3\epsilon\cH^2)^{-n}$,
which mix the wavelength scale $k^2$
with the slow-roll geometrical scale $\epsilon\cH^2$.
A series expansion of these factors generates
an infinite sequence of terms,
and expanding first in wavelength
or first in slow roll gives different series
corresponding to different asymptotic hierarchies.
The UE series expansion therefore cannot be regarded simply
as a finite reordering of terms
already present in the original 2EMT.
\subsection{Definition of the 2EMT}

At second perturbative order,
the Einstein equation can be arranged as
\begin{equation}
 G^{(1)}_{\mu\nu}[g^{(2)}]
 =8\pi G T^{(1)}_{\mu\nu}[g^{(2)},\delta\phi^{(2)}]
 +8\pi G T^{(2,\mathrm{eff})}_{\mu\nu},
\end{equation}
where
\begin{equation}
 T^{(2,\mathrm{eff})}_{\mu\nu}
 \equiv T^{(2)}_{\mu\nu}[g^{(1)},\delta\phi^{(1)}]
 -\frac{1}{8\pi G}G^{(2)}_{\mu\nu}[g^{(1)}].
\label{eq:2EMTdef}
\end{equation}
The quadratic effective source is evaluated in Fourier space.
We denote the contribution associated
with a fixed Fourier pair $\bm k$ and $-\bm k$ by
\begin{equation}
 \widehat\tau_{\mu\nu}(k)
 \equiv8\pi G\tau_{\mu\nu}(k)
 =8\pi G\left\langle T^{(2,\mathrm{eff})}_{\mu\nu}\right\rangle_k.
\label{eq:tauaverage}
\end{equation}
Here, $\langle\cdots\rangle_k$ denotes
the Fourier-transformed quadratic quantity
obtained through the required spatial integration;
it does not represent a separate averaging prescription.
The integration pairs the modes $\bm k$ and $-\bm k$.
For a real perturbation,
$\Psi_{-\bm k}=\Psi_{\bm k}^{*}$,
and similarly for the other fields.
Below we suppress the explicit argument $k$
of the fixed-mode contribution.

The fixed Fourier pair is not itself isotropic.
For the projected scalar sector we use,
following Abramo et al.~\cite{Abramo:1997hu},
\begin{equation}
 k_i k_j\longrightarrow\frac{k^2}{3}\delta_{ij},
 \qquad
 \widehat\tau_{ii}\equiv\frac13\delta^{ij}\widehat\tau_{ij}.
\label{eq:isoproj}
\end{equation}
Thus $\widehat\tau_{ii}$ denotes one third
of the projected spatial trace,
and the off-diagonal spatial components vanish after this projection.
The corresponding diagnostic energy density,
pressure, and ratio are
\begin{equation}
 \rho_{\rm raw}=\frac{\tau_{00}}{a^2},
 \qquad
 p_{\rm raw}=\frac{\tau_{ii}}{a^2},
 \qquad
 w_{\rm raw}=\frac{p_{\rm raw}}{\rho_{\rm raw}}.
\label{eq:wraw}
\end{equation}
The label ``raw'' emphasizes
that these gauge-fixed quantities are not,
by themselves, relational observables.
In particular, $w_{\rm raw}$ is introduced only
as a diagnostic component ratio
and should not be identified
with a physical equation-of-state parameter
without a common observer and relational prescription.

The general Fourier-space expression for $\widehat\tau_{\mu\nu}$
depends on the gauge-invariant potential $\Psi$
and on the gauge variables $Q$ and $E$
\cite{Cho:2022maa,Cho:2026gjy}.
In the longitudinal, spatially-flat, comoving,
and uniform-density gauges,
the imposed conditions allow the final reduced expressions
to be written directly in terms of $\Psi$.
The uniform-expansion gauge requires one additional structural step:
in configuration space its temporal condition
is a \textit{differential} equation for $Q$,
while in Fourier space that equation becomes
an \textit{algebraic} relation for each Fourier mode.
\section{Uniform-expansion gauge, gauge dependence, and complete gauge fixing}
\label{sec:UE}

\subsection{Gauge transformations, gauge invariance, and the general quadratic source}
\label{subsec:rawGT}

We first recall the meaning of gauge transformation and gauge
invariance in cosmological perturbation theory. A gauge transformation
does not represent a physical change of the perturbed spacetime.
Rather, it changes the identification map that associates points of
the homogeneous background spacetime with points of the physical
spacetime when perturbations are defined. An identification map may
be written as
\begin{equation}
 X:\mathcal M_0\to\mathcal M,
\end{equation}
where $\mathcal M_0$ denotes the homogeneous background spacetime
and $\mathcal M$ the physical perturbed spacetime.

A change of identification map from $X$ to $Y$ generally associates
the same point of the background spacetime with different points of
the physical spacetime. This change gives rise to the gauge
transformation laws of the perturbation variables. The physical
spacetime and its fields remain unchanged; what changes is their
identification with the background spacetime. For a fixed physical
spacetime, the admissible identification maps
$X,Y,\ldots:\mathcal M_0\to\mathcal M$ generate gauge-related
perturbative configurations, which form a gauge orbit.

At first order, the scalar part of a change of identification map
is generated by
\begin{equation}
 \xi^\mu=(\xi^0,\partial^i\zeta),
\end{equation}
and the corresponding transformations of $\alpha$, $\beta$, $\psi$,
$E$, and $\delta\phi$ are given in Eqs.~(7)--(11). In particular,
$Q$ and $E$ transform as
$\widetilde Q=Q-\xi^0$ and $\widetilde E=E-\zeta$, whereas the
Bardeen potentials $\Phi$ and $\Psi$ remain gauge invariant.
This is the standard scalar-perturbation notion of gauge freedom
\cite{Mukhanov:1990me,Malik:2008im}.

Writing
\begin{equation}
 \mathcal X=\{\alpha,\beta,\psi,E,Q,\delta\phi,\ldots\},
\end{equation}
a perturbative functional $F[\mathcal X]$ is gauge invariant if it
has the same value for arbitrary gauge-related configurations,
\begin{equation}
 F[\widetilde{\mathcal X}]=F[\mathcal X]
 \qquad
 \text{for arbitrary }\xi^0\text{ and }\zeta,
\label{eq:GIcriterion}
\end{equation}
or equivalently
\begin{equation}
 \Delta_\xi F
 \equiv
 F[\widetilde{\mathcal X}]-F[\mathcal X]=0.
\label{eq:deltaFcriterion}
\end{equation}
Equation~\eqref{eq:deltaFcriterion} is the gauge-invariance
criterion before any gauge condition is imposed.

For the quadratic effective source, this criterion must be applied
to the complete general 2EMT. The presence of individually
gauge-dependent variables such as $Q$ and $E$ does not by itself
establish gauge dependence, since their transformed contributions
could in principle cancel in the complete quadratic expression.
The decisive test is therefore to apply an arbitrary scalar gauge
transformation to the general quadratic source before gauge fixing,
and evaluate its variation directly.

The gauge status of the quadratic effective source has been a
long-standing issue in the backreaction literature. The early
2EMT program formulated the quadratic source with the aim of
obtaining a gauge-invariant description
\cite{Mukhanov:1996ak,Abramo:1997hu}, whereas later analyses
raised concerns about gauge dependence in the 2EMT framework
\cite{Unruh:1998ic,Ishibashi:2005sj}.
Our previous studies evaluated the scalar 2EMT in different
gauge choices
\cite{Cho:2020zbh,Cho:2022maa,Cho:2024vpv,Cho:2026gjy}.
Here, we perform the direct transformation test before imposing
any temporal or spatial gauge condition. To our knowledge, this
is the first explicit component-level evaluation of the general
scalar-gauge variation of the quadratic source in the present
2EMT framework.

We start from the general scalar quadratic source
$T_{\mu\nu}^{(2,\mathrm{eff})}$ given in Appendix~A of
Ref.~\cite{Cho:2026gjy}. Defining
\begin{equation}
 \mathcal S_{\mu\nu}[\mathcal X]
 \equiv
 8\pi G T_{\mu\nu}^{(2,\mathrm{eff})}[\mathcal X],
\end{equation}
we consider its variation in configuration space,
\begin{equation}
 \Delta_\xi\mathcal S_{\mu\nu}
 \equiv
 \mathcal S_{\mu\nu}[\widetilde{\mathcal X}]
 -
 \mathcal S_{\mu\nu}[\mathcal X].
\label{eq:rawGTdef}
\end{equation}
No gauge condition is imposed, and the transformation parameters
$\xi^0$ and $\zeta$ remain arbitrary.

The explicit results for the $00$ and $ii$ components, with no sum
over $i$, are given in Appendix~\ref{app:rawGT}. In simplifying
these expressions, terms that are odd under reversal of the
integration momentum, including the total-spatial-derivative
contributions encountered in the calculation, are omitted. The
remaining expressions contain nonvanishing even-in-$k$
contributions. Since these even and odd sectors cannot cancel
identically, we obtain
\begin{equation}
 \Delta_\xi\mathcal S_{00}\neq0,
 \qquad
 \Delta_\xi\mathcal S_{ii}\neq0
 \qquad
 \text{in general (no sum over }i\text{)}.
\label{eq:rawGTnonzero}
\end{equation}
Thus, the general scalar 2EMT before gauge fixing does not satisfy
the gauge-invariance criterion in Eq.~\eqref{eq:GIcriterion}.
The general quadratic source itself is therefore
\textit{gauge dependent} under an arbitrary scalar gauge
transformation.

The spatial off-diagonal components require a separate technical
point. After the terms odd in the integration momentum are removed
under symmetric momentum integration, the remaining off-diagonal
tensor structures proportional to $k_i k_j$ with $i\neq j$ vanish
after isotropic angular integration. Consequently, the scalar
spatial source obtained in the Fourier calculation is diagonal,
\begin{equation}
 \widehat\tau_{ij}(k)
 \longrightarrow
 \delta_{ij}\widehat\tau_s(k).
\end{equation}
The same reduction applies to the corresponding gauge-variation
terms. These cancellations are part of the Fourier integration
used in constructing the quadratic source and do not constitute
an additional spatial-averaging prescription.

The discussion above concerns the general quadratic source before
gauge fixing. Gauge fixing addresses a different question: it
selects an identification map, and hence a perturbative
representation, by imposing specified conditions on the
perturbation variables. We now turn to the uniform-expansion
choice.

\subsection{Uniform-expansion condition}
\label{subsec:UEcondition}

To make the geometrical meaning of the gauge explicit, let $n^\mu$
be the future-directed unit normal to the constant-$\eta$
hypersurfaces, and let
\begin{equation}
 h_{\mu\nu}=g_{\mu\nu}+n_\mu n_\nu
\end{equation}
be the induced metric on these hypersurfaces.

We adopt the following sign convention for the extrinsic curvature:
\begin{equation}
 \mathcal K_{\mu\nu}
 =-h_\mu{}^\rho h_\nu{}^\sigma\nabla_\rho n_\sigma.
\end{equation}
Then the expansion scalar of the normal congruence is
\begin{equation}
 \Theta\equiv\nabla_\mu n^\mu=-\mathcal K,
 \qquad
 \mathcal K\equiv\mathcal K^\mu{}_{\mu}.
\label{eq:ThetaK}
\end{equation}

For the scalar metric of Eq.~\eqref{eq:metric}, the linear spatial
extrinsic curvature is
\begin{equation}
 \mathcal K^i{}_{j}
 =-\frac{\cH}{a}\delta^i{}_j
 +\frac{1}{a}
 \left[
  (\psi'+\cH\alpha)\delta^i{}_j
  -Q^{,i}{}_{,j}
 \right],
\label{eq:Kij}
\end{equation}
and its trace is conventionally written as
\begin{equation}
 \mathcal K=-\frac{3\cH}{a}+\frac{\kappa}{a},
 \qquad
 \kappa\equiv3(\psi'+\cH\alpha)-\Delta Q.
\label{eq:kappadef}
\end{equation}
Equivalently, at first perturbative order,
\begin{equation}
 \Theta=\frac{3\cH}{a}-\frac{\kappa}{a},
 \qquad
 \delta\Theta^{(1)}=-\frac{\kappa}{a}.
\label{eq:ThetaKappa}
\end{equation}

The temporal uniform-expansion condition requires the first-order
perturbation of the expansion scalar to vanish, while the remaining
scalar spatial gauge freedom is fixed independently by $E=0$.
The uniform-expansion gauge is therefore
\begin{equation}
 \delta\Theta^{(1)}=0\quad(\kappa=0),
 \qquad
 E=0.
\label{eq:UEgauge}
\end{equation}

Cosmological backreaction concerns how inhomogeneous perturbations
contribute to the effective evolution of a homogeneous background,
particularly its expansion. The uniform-expansion gauge provides
a geometrically defined temporal slicing on which the first-order
perturbation of the expansion scalar of the hypersurface-normal
congruence vanishes. This gauge condition does not eliminate the
perturbations themselves or their quadratic contributions. Rather,
it specifies the hypersurfaces on which the 2EMT is evaluated,
allowing us to investigate how the quadratic effective source
depends on a slicing defined by uniform local expansion.

At first order, the uniform-expansion condition
$\delta\Theta^{(1)}=0$ ($\kappa=0$) corresponds to
constant-mean-curvature (CMC) slicing, since the trace of the
extrinsic curvature is spatially uniform on each constant-$\eta$
hypersurface, although it may evolve in time.
The variable $\kappa$ and the extrinsic-curvature trace are
standard in gauge-ready formulations of cosmological perturbation
theory
\cite{Harada:2015yda,Mukhanov:1990me,Malik:2008im}.

From the definitions of $\Phi$ and $\Psi$, together with
$\Phi=\Psi$, we have
\begin{equation}
 \psi=\Psi-\cH Q,
 \qquad
 \alpha=\Psi+Q'+\cH Q.
\end{equation}
Then Eq.~\eqref{eq:kappadef} becomes
\begin{equation}
 \kappa
 =3(\Psi'+\cH\Psi)
 +\left[3(\cH^2-\cH')-\Delta\right]Q.
\label{eq:kappaPsiQ}
\end{equation}
The condition $\kappa=0$ therefore gives
\begin{equation}
 (\Delta-3\epsilon\cH^2)Q_{\ue}
 =3(\Psi'+\cH\Psi),
\label{eq:Qconstraintx}
\end{equation}
where Eq.~\eqref{eq:Hprime} has been used.

Equation~\eqref{eq:Qconstraintx} is the first structural
distinction from the gauges considered in the companion analysis.
In configuration space, it is a \textit{differential} equation
for $Q$, rather than an \textit{algebraic} relation.
Consequently, an unreduced derivative representation of the 2EMT
can retain $Q$, $Q'$, and $Q''$; the explicit expressions are
collected in Appendix~\ref{app:unreduced}.

For a Fourier mode, $\Delta\to-k^2$, and
Eq.~\eqref{eq:Qconstraintx} becomes
\begin{equation}
 Q_{\ue}(k)
 =-\frac{3(\Psi'+\cH\Psi)}
 {k^2+3\epsilon\cH^2}.
\label{eq:QUE}
\end{equation}
The remaining perturbations are therefore
\begin{align}
 E_{\ue}&=0,\qquad \beta_{\ue}=Q_{\ue},\\
 \psi_{\ue}&=\Psi-\cH Q_{\ue},\\
 \alpha_{\ue}&=\Psi+Q_{\ue}'+\cH Q_{\ue},\\
 \delta\phi_{\ue}
 &=\delta\phi_{\rm gi}+\phi_0'Q_{\ue}.
\end{align}

Equation~\eqref{eq:QUE} immediately exposes the scale competition
in Eq.~\eqref{eq:charfactor}. It also gives useful limiting checks.
Since
\begin{equation}
 Q_{\cgauge}
 =\frac{\Psi'+\cH\Psi}{\cH'-\cH^2}
 =-\frac{\Psi'+\cH\Psi}{\epsilon\cH^2},
\end{equation}
we have
\begin{equation}
 Q_{\ue}\longrightarrow Q_{\cgauge}
 \qquad
 (k^2\ll\epsilon\cH^2).
\end{equation}
For $k^2\gg\epsilon\cH^2$, by contrast,
$Q_{\ue}=O(k^{-2})$, and the shift becomes gradient suppressed,
which anticipates the approach to the spatially-flat result.

\subsection{Complete gauge fixing and the gauge-fixed 2EMT}
\label{subsec:completion}

Gauge fixing selects, within a given gauge orbit, an identification
map for which the specified gauge conditions are satisfied.
Starting from an arbitrary identification map, the gauge conditions
determine the transformation parameters required to reach the
selected representative. A complete gauge fixing removes any
residual gauge freedom and therefore makes this representative
unique, subject to the appropriate boundary or regularity
conditions.

For the uniform-expansion gauge, the transformation laws
\begin{equation}
 \widetilde\kappa
 =\kappa+(\Delta-3\epsilon\cH^2)\xi^0,
 \qquad
 \widetilde E=E-\zeta
\label{eq:kappatrans}
\end{equation}
show directly that the fixing is complete. If both the original
and transformed configurations satisfy the same gauge conditions,
$\kappa=\widetilde\kappa=0$ and $E=\widetilde E=0$, then
$\zeta=0$ and
\begin{equation}
 (\Delta-3\epsilon\cH^2)\xi^0=0.
\end{equation}
With the boundary or regularity conditions that remove the
homogeneous solution, $\xi^0=0$, and no nontrivial residual scalar
gauge transformation remains.

Once the complete gauge has been fixed, the perturbation variables
and the corresponding 2EMT are evaluated within this selected
representation. The absence of residual gauge freedom means that
the gauge-fixed result is unambiguously specified for a given
perturbative solution.

In particular, the \emph{variables} appearing in the final gauge-fixed
expression do not determine its gauge status. For example, in the
spatially-flat gauge $\psi_{\rm SF}=0$, and hence
$\Psi=\cH Q_{\rm SF}$. A gauge-fixed 2EMT written in terms of
$\Psi$ may therefore equally be written in terms of $Q_{\rm SF}$.
The two expressions represent the same quantity in the same fixed
gauge. Thus, the appearance of a gauge-invariant variable such as
$\Psi$, or of a gauge variable such as $Q$,
does not by itself implement gauge invariance or imply gauge
dependence of the gauge-fixed 2EMT.

In the following sections, the uniform-expansion conditions are
maintained throughout the evaluation and interpretation of the
corresponding quadratic source.

\section{Uniform-expansion 2EMT in the long-wavelength domain}
\label{sec:UELW}

We substitute the UE constraint
and the long-wavelength mode
into the general scalar 2EMT.
We use
$\sigma_1=k/\cH\ll1$.
The exact result is rational
in $k^2+3\epsilon\cH^2$.
The second representation below
is the wavelength-first expansion
for the strict infrared.
The third is the slow-roll-first expansion
for the intermediate infrared.\footnote{We have checked whether the additional
$\Order(k^2)$ correction to the mode $\Psi_k$ in {\it long-wavelength limit}
in Eq.~\eqref{eq:PsiSolutions} affects the first nonzero infrared
terms of the UE 2EMT. 
Using the linear mode equation Eq.~\eqref{eq:PsiEOM} 
and the exact background identities in Eq.~\eqref{eq:bgconsistent}, 
we have an overall $k^2$ factor in both $\widehat\tau_{00}^{\mathrm{LW}}$
and $\widehat\tau_{ii}^{\mathrm{LW}}$
even before specifying $\Psi_k$ and $\Psi_k'$.
Consequently, in the strict IR, the mode correction first contributes
at $\Order(k^4)$ and leaves the first nonzero $k^2$ coefficients unchanged.
This holds for general slow-roll backgrounds.
A test calculation on a constant-$\epsilon$ background shows that
the mode correction changes the $\sigma_1^4$ coefficient
in the slow-roll-first intermediate-IR expansion.}

For comparison with the other gauges,
we define
\begin{equation}
\factLW\equiv
\cH^2|A_1|^2.
\label{eq:FLWdef}
\end{equation}

\begin{align}
\widehat\tau_{00}^{\mathrm{LW}}={}&
\frac{k^2\epsilon|A_1|^2}{\cH^2(k^2+3\epsilon\cH^2)^4}
\Bigg\{k^{10}+\cH^2k^8[1-2\delta+\epsilon(29-4\delta)+4\epsilon^2]
+\cH^4k^6[\delta(1-2\delta)^2(\delta+3)+24\epsilon^3
\nonumber\\
&\hspace{18mm}
+\epsilon^2(4\delta^2+40\delta+250)-2\epsilon(4\delta^3+36\delta^2-\delta+39)]
+6\cH^6\epsilon k^4[12\delta^4-8\delta^3+107\delta^2-95\delta+21
\nonumber\\
&\hspace{18mm}
+3\epsilon^2(4\delta^2+26\delta+61)-\epsilon(24\delta^3+70\delta^2+135\delta+44)]
-9\cH^8\epsilon k^2[24\epsilon^4-3\epsilon^3(12\delta^2+32\delta+111)
\nonumber\\
&\hspace{18mm}
+6\epsilon^2(12\delta^3+8\delta^2+95\delta-9)-\epsilon(36\delta^4+24\delta^3+349\delta^2-361\delta+88)
+12(1-2\delta)^2(\delta^2-1)]
\nonumber\\
&\hspace{18mm}
-81\cH^{10}\epsilon^2[4\epsilon^4-\epsilon^3(8\delta+41)+\epsilon^2(4\delta^2+108\delta-31)
-2\epsilon(52\delta^2-56\delta+15)+8(1-2\delta)^2(\delta-1)]
\nonumber\\
&\hspace{18mm}
-\cH^4k^2\xi_3(k^2+3\epsilon\cH^2)^2(1+2\epsilon-2\delta)^2\Bigg\}\nonumber\\[1mm] \simeq{}& \factLW \Bigg\{ \frac{\sigma_1^{2}}{\epsilon}\Big[ 8 +\bigl(- 40 \delta + 30 \epsilon\bigr) +\bigl(64 \delta^{2} - 112 \delta \epsilon + 31 \epsilon^{2}\bigr)
\nonumber\\
&\hspace{18mm}
+\bigl(- 32 \delta^{3} + 104 \delta^{2} \epsilon - 108 \delta \epsilon^{2} + 41 \epsilon^{3}\bigr) +\bigl(- 4 \delta^{2} \epsilon^{2} + 8 \delta \epsilon^{3} - 4 \epsilon^{4}\bigr) \Big]\nonumber\\
&\hspace{13mm}
+ \frac{\sigma_1^{4}}{9 \epsilon^{2}}\Big[ -84 +\bigl(432 \delta - 272 \epsilon\bigr) +\bigl(- 732 \delta^{2} + 983 \delta \epsilon - 318 \epsilon^{2}\bigr)
\nonumber\\
&\hspace{18mm}
+\bigl(432 \delta^{3} - 899 \delta^{2} \epsilon + 726 \delta \epsilon^{2} - 159 \epsilon^{3} - \epsilon \xi_3\bigr)
\nonumber\\
&\hspace{18mm}
+\bigl(- 48 \delta^{4} + 24 \delta^{3} \epsilon + 4 \delta \epsilon \xi_3 + 24 \epsilon^{4} - 4 \epsilon^{2} \xi_3\bigr)
\nonumber\\
&\hspace{18mm}
+\bigl(36 \delta^{4} \epsilon - 72 \delta^{3} \epsilon^{2} + 36 \delta^{2} \epsilon^{3} - 4 \delta^{2} \epsilon \xi_3 + 8 \delta \epsilon^{2} \xi_3 - 4 \epsilon^{3} \xi_3\bigr) \Big]\nonumber\\
&\hspace{13mm}
+ \frac{\sigma_1^{6}}{27 \epsilon^{3}}\Big[ 192 +\bigl(- 1008 \delta + 590 \epsilon\bigr) +\bigl(1776 \delta^{2} - 2106 \delta \epsilon + 626 \epsilon^{2}\bigr)
\nonumber\\
&\hspace{18mm}
+\bigl(- 1152 \delta^{3} + 1938 \delta^{2} \epsilon - 1230 \delta \epsilon^{2} + 264 \epsilon^{3} + 2 \epsilon \xi_3\bigr)
\nonumber\\
&\hspace{18mm}
+\bigl(192 \delta^{4} - 112 \delta^{3} \epsilon - 68 \delta^{2} \epsilon^{2} + 12 \delta \epsilon^{3} - 8 \delta \epsilon \xi_3 - 24 \epsilon^{4} + 8 \epsilon^{2} \xi_3\bigr)
\nonumber\\
&\hspace{18mm}
+\bigl(- 120 \delta^{4} \epsilon + 240 \delta^{3} \epsilon^{2} - 120 \delta^{2} \epsilon^{3} + 8 \delta^{2} \epsilon \xi_3 - 16 \delta \epsilon^{2} \xi_3 + 8 \epsilon^{3} \xi_3\bigr) \Big]\nonumber\\
&\hspace{13mm}
+ \frac{\sigma_1^{8}}{81 \epsilon^{4}}\Big[ -360 +\bigl(1920 \delta - 1088 \epsilon\bigr) +\bigl(- 3480 \delta^{2} + 3873 \delta \epsilon - 1046 \epsilon^{2}\bigr)
\nonumber\\
&\hspace{18mm}
+\bigl(2400 \delta^{3} - 3617 \delta^{2} \epsilon + 1862 \delta \epsilon^{2} - 344 \epsilon^{3} - 3 \epsilon \xi_3\bigr)
\nonumber\\
&\hspace{18mm}
+\bigl(- 480 \delta^{4} + 312 \delta^{3} \epsilon + 248 \delta^{2} \epsilon^{2} - 104 \delta \epsilon^{3} + 12 \delta \epsilon \xi_3 + 24 \epsilon^{4} - 12 \epsilon^{2} \xi_3\bigr)
\nonumber\\
&\hspace{18mm}
+\bigl(268 \delta^{4} \epsilon - 536 \delta^{3} \epsilon^{2} + 268 \delta^{2} \epsilon^{3} - 12 \delta^{2} \epsilon \xi_3 + 24 \delta \epsilon^{2} \xi_3 - 12 \epsilon^{3} \xi_3\bigr) \Big]\nonumber\\
&\hspace{13mm}
+ \frac{\sigma_1^{10}}{243 \epsilon^{5}}\Big[ 600 +\bigl(- 3240 \delta + 1810 \epsilon\bigr) +\bigl(6000 \delta^{2} - 6452 \delta \epsilon + 1610 \epsilon^{2}\bigr)
\nonumber\\
&\hspace{18mm}
+\bigl(- 4320 \delta^{3} + 6124 \delta^{2} \epsilon - 2654 \delta \epsilon^{2} + 392 \epsilon^{3} + 4 \epsilon \xi_3\bigr)
\nonumber\\
&\hspace{18mm}
+\bigl(960 \delta^{4} - 672 \delta^{3} \epsilon - 572 \delta^{2} \epsilon^{2} + 308 \delta \epsilon^{3} - 16 \delta \epsilon \xi_3 - 24 \epsilon^{4} + 16 \epsilon^{2} \xi_3\bigr)
\nonumber\\
&\hspace{18mm}
+\bigl(- 496 \delta^{4} \epsilon + 992 \delta^{3} \epsilon^{2} - 496 \delta^{2} \epsilon^{3} + 16 \delta^{2} \epsilon \xi_3
- 32 \delta \epsilon^{2} \xi_3 + 16 \epsilon^{3} \xi_3\bigr) \Big]
\nonumber\\
&\hspace{13mm}+\Order(\sigma_1^{12}) \Bigg\} \qquad\text{(strict IR)} \nonumber\\[1mm]\simeq{}& \epsilon\factLW \Bigg\{ \left[ \sigma_1^2+\sigma_1^4 \right]\nonumber\\
&\hspace{13mm}
+\left[
3\delta-90\epsilon
+\frac{126\epsilon}{\sigma_1^2}
+\frac{108\epsilon}{\sigma_1^4}
+(-2\delta+17\epsilon)\sigma_1^2
\right]
+\Order_{\rm SR}(2;\sigma_1)
\Bigg\}
\qquad\text{(intermediate IR)}.
\label{eq:UElw00full}
\end{align}

\begin{align}
\widehat\tau_{ii}^{\mathrm{LW}}={}&
\frac{k^2\epsilon|A_1|^2}{3\cH^2(k^2+3\epsilon\cH^2)^5}
\Bigg\{3k^{12}-\cH^2k^{10}[1-10\delta+16\delta^2+4\epsilon^2-2\epsilon(10\delta+23)]
\nonumber\\
&\hspace{18mm}
-3\cH^4k^8[-\delta(1-2\delta)^2(\delta-3)+28\epsilon^3-\epsilon^2(4\delta^2+84\delta+239)
+\epsilon(8\delta^3+40\delta^2+228\delta-145)]
\nonumber\\
&\hspace{18mm}
-9\cH^6\epsilon k^6[4\delta^4-272\delta^3+597\delta^2-323\delta+46+48\epsilon^3
+4\epsilon^2(\delta^2-67\delta-111)-4\epsilon(2\delta^3-123\delta^2+13\delta+44)]
\nonumber\\
&\hspace{18mm}
-27\cH^8\epsilon k^4[16\epsilon^4+\epsilon^3(36\delta^2-356\delta-367)-4\epsilon^2(10\delta^3-141\delta^2+94\delta+20)
\nonumber\\
&\hspace{18mm}
+\epsilon(-28\delta^4-80\delta^3+629\delta^2-447\delta+78)+4(1-2\delta)^2(2\delta^3-7\delta^2-4\delta+7)]
\nonumber\\
&\hspace{18mm}
+81\cH^{10}\epsilon^2k^2[20\epsilon^4+\epsilon^3(-28\delta^2+80\delta+230)+\epsilon^2(152\delta^3-244\delta^2+274\delta-39)
\nonumber\\
&\hspace{18mm}
+\epsilon(-220\delta^4+416\delta^3-915\delta^2+745\delta-182)+4(1-2\delta)^2(6\delta^3-11\delta^2+26\delta-17)]
\nonumber\\
&\hspace{18mm}
+243\cH^{12}\epsilon^3[12\epsilon^4+\epsilon^3(105-72\delta)+\epsilon^2(236\delta^2-196\delta+39)
-2\epsilon(136\delta^3-128\delta^2+8\delta+11)
\nonumber\\
&\hspace{18mm}
+8(1-2\delta)^2(3\delta^2-\delta-1)]
\nonumber\\
&\hspace{18mm}
+3\cH^4\xi_3(k^2+3\epsilon\cH^2)[-216\cH^6\epsilon^2\{6\epsilon^2+\epsilon(5-10\delta)+(1-2\delta)^2\}
+9\cH^4\epsilon k^2\{-8\delta(1-2\delta)^2
\nonumber\\
&\hspace{18mm}
+4\epsilon^3-20\epsilon^2(2\delta+3)+\epsilon(68\delta^2+28\delta-31)\}
+6\cH^2\epsilon k^4\{4\delta^2+20\delta+4\epsilon^2-4\epsilon(2\delta+5)-11\}
\nonumber\\
&\hspace{18mm}
+k^6(1+2\epsilon-2\delta)^2]\Bigg\}\nonumber\\[1mm] \simeq{}& \factLW \Bigg\{ \frac{\sigma_1^{2}}{3 \epsilon}\Big[ -8 +\bigl(24 \delta - 22 \epsilon\bigr) +\bigl(24 \delta^{2} - 16 \delta \epsilon + 39 \epsilon^{2} - 8 \xi_3\bigr)
\nonumber\\
&\hspace{18mm}
+\bigl(- 128 \delta^{3} + 256 \delta^{2} \epsilon - 196 \delta \epsilon^{2} + 32 \delta \xi_3 + 105 \epsilon^{3} - 40 \epsilon \xi_3\bigr)
\nonumber\\
&\hspace{18mm}
+\bigl(96 \delta^{4} - 272 \delta^{3} \epsilon + 236 \delta^{2} \epsilon^{2} - 32 \delta^{2} \xi_3 - 72 \delta \epsilon^{3} + 80 \delta \epsilon \xi_3 + 12 \epsilon^{4} - 48 \epsilon^{2} \xi_3\bigr) \Big]\nonumber\\
&\hspace{13mm}
+ \frac{\sigma_1^{4}}{9 \epsilon^{2}}\Big[ -28 +\bigl(256 \delta - 72 \epsilon\bigr) +\bigl(- 852 \delta^{2} + 825 \delta \epsilon - 234 \epsilon^{2} + 32 \xi_3\bigr)
\nonumber\\
&\hspace{18mm}
+\bigl(1256 \delta^{3} - 2195 \delta^{2} \epsilon + 1254 \delta \epsilon^{2} - 136 \delta \xi_3 - 295 \epsilon^{3} + 129 \epsilon \xi_3\bigr)
\nonumber\\
&\hspace{18mm}
+\bigl(- 752 \delta^{4} + 1776 \delta^{3} \epsilon - 1424 \delta^{2} \epsilon^{2} + 160 \delta^{2} \xi_3 + 440 \delta \epsilon^{3} - 292 \delta \epsilon \xi_3 - 40 \epsilon^{4} + 132 \epsilon^{2} \xi_3\bigr)
\nonumber\\
&\hspace{18mm}
+\bigl(96 \delta^{5} - 220 \delta^{4} \epsilon + 152 \delta^{3} \epsilon^{2} - 32 \delta^{3} \xi_3 - 28 \delta^{2} \epsilon^{3} + 68 \delta^{2} \epsilon \xi_3 - 40 \delta \epsilon^{2} \xi_3 + 4 \epsilon^{3} \xi_3\bigr) \Big]\nonumber\\
&\hspace{13mm}
+ \frac{\sigma_1^{6}}{27 \epsilon^{3}}\Big[ 192 +\bigl(- 1392 \delta + 502 \epsilon\bigr) +\bigl(3872 \delta^{2} - 3518 \delta \epsilon + 860 \epsilon^{2} - 80 \xi_3\bigr)
\nonumber\\
&\hspace{18mm}
+\bigl(- 5056 \delta^{3} + 7786 \delta^{2} \epsilon - 3934 \delta \epsilon^{2} + 352 \delta \xi_3 + 792 \epsilon^{3} - 298 \epsilon \xi_3\bigr)
\nonumber\\
&\hspace{18mm}
+\bigl(2944 \delta^{4} - 6080 \delta^{3} \epsilon + 4196 \delta^{2} \epsilon^{2} - 448 \delta^{2} \xi_3 - 1124 \delta \epsilon^{3} + 728 \delta \epsilon \xi_3 + 64 \epsilon^{4} - 280 \epsilon^{2} \xi_3\bigr)
\nonumber\\
&\hspace{18mm}
+\bigl(- 512 \delta^{5} + 1128 \delta^{4} \epsilon - 720 \delta^{3} \epsilon^{2} + 128 \delta^{3} \xi_3 + 104 \delta^{2} \epsilon^{3} - 264 \delta^{2} \epsilon \xi_3 + 144 \delta \epsilon^{2} \xi_3 - 8 \epsilon^{3} \xi_3\bigr) \Big]\nonumber\\
&\hspace{13mm}
+ \frac{\sigma_1^{8}}{81 \epsilon^{4}}\Big[ -600 +\bigl(4160 \delta - 1616 \epsilon\bigr) +\bigl(- 11080 \delta^{2} + 9823 \delta \epsilon - 2174 \epsilon^{2} + 160 \xi_3\bigr)
\nonumber\\
&\hspace{18mm}
+\bigl(14000 \delta^{3} - 20137 \delta^{2} \epsilon + 9142 \delta \epsilon^{2} - 720 \delta \xi_3 - 1616 \epsilon^{3} + 579 \epsilon \xi_3\bigr)
\nonumber\\
&\hspace{18mm}
+\bigl(- 8160 \delta^{4} + 15632 \delta^{3} \epsilon - 9592 \delta^{2} \epsilon^{2} + 960 \delta^{2} \xi_3 + 2208 \delta \epsilon^{3} - 1484 \delta \epsilon \xi_3 - 88 \epsilon^{4} + 524 \epsilon^{2} \xi_3\bigr)
\nonumber\\
&\hspace{18mm}
+\bigl(1600 \delta^{5} - 3444 \delta^{4} \epsilon + 2088 \delta^{3} \epsilon^{2} - 320 \delta^{3} \xi_3 - 244 \delta^{2} \epsilon^{3} + 652 \delta^{2} \epsilon \xi_3 - 344 \delta \epsilon^{2} \xi_3 + 12 \epsilon^{3} \xi_3\bigr) \Big]\nonumber\\
&\hspace{13mm}
+ \frac{\sigma_1^{10}}{243 \epsilon^{5}}\Big[ 1400 +\bigl(- 9560 \delta + 3890 \epsilon\bigr) +\bigl(25080 \delta^{2} - 22108 \delta \epsilon + 4560 \epsilon^{2} - 280 \xi_3\bigr)
\nonumber\\
&\hspace{18mm}
+\bigl(- 31360 \delta^{3} + 43508 \delta^{2} \epsilon - 18158 \delta \epsilon^{2} + 1280 \delta \xi_3 + 2824 \epsilon^{3} - 1004 \epsilon \xi_3\bigr)
\nonumber\\
&\hspace{18mm}
+\bigl(18400 \delta^{4} - 33776 \delta^{3} \epsilon + 19020 \delta^{2} \epsilon^{2} - 1760 \delta^{2} \xi_3 - 3756 \delta \epsilon^{3} + 2656 \delta \epsilon \xi_3 + 112 \epsilon^{4} - 896 \epsilon^{2} \xi_3\bigr)
\nonumber\\
&\hspace{18mm}
+\bigl(- 3840 \delta^{5} + 8144 \delta^{4} \epsilon - 4768 \delta^{3} \epsilon^{2} + 640 \delta^{3} \xi_3 + 464 \delta^{2} \epsilon^{3} - 1296 \delta^{2} \epsilon \xi_3 + 672 \delta \epsilon^{2} \xi_3 - 16 \epsilon^{3} \xi_3\bigr) \Big]
\nonumber\\
&\hspace{13mm}+\Order(\sigma_1^{12}) \Bigg\} \qquad\text{(strict IR)} \nonumber\\[1mm]\simeq{}& \epsilon\factLW \Bigg\{ \left[ -\frac13\sigma_1^2+\sigma_1^4 \right]\nonumber\\
&\hspace{13mm}
+\left[
-3\delta+150\epsilon
-\frac{138\epsilon}{\sigma_1^2}
-\frac{252\epsilon}{\sigma_1^4}
+\frac{10\delta+\epsilon}{3}\sigma_1^2
\right]
+\Order_{\rm SR}(2;\sigma_1)
\Bigg\}
\qquad\text{(intermediate IR)}.
\label{eq:UElwiifull}
\end{align}

\subsection{Strict infrared}

The strict infrared is defined
by $\sigma_1\to0$
at fixed admissible nonzero
slow-roll parameters.
Neither component contains
a $\sigma_1^0$ term.
Therefore
\begin{equation}
\widehat\tau_{00}^{\ue},
\widehat\tau_{ii}^{\ue}
\longrightarrow0
\qquad
(\sigma_1\to0).
\label{eq:UEzeroIR}
\end{equation}

The first nonzero terms are
\begin{align}
\frac{\widehat\tau_{00}^{\ue}}{\factLW}
={}&
\frac{\sigma_1^2}{\epsilon}
\Big[
8+(30\epsilon-40\delta)
+\Order_{\rm SR}(2)
\Big]
+\Order(\sigma_1^4),
\nonumber\\
\frac{\widehat\tau_{ii}^{\ue}}{\factLW}
={}&
-\frac{\sigma_1^2}{3\epsilon}
\Big[
8+(22\epsilon-24\delta)
+\Order_{\rm SR}(2)
\Big]
+\Order(\sigma_1^4).
\label{eq:UEstrictIRfirst}
\end{align}
Hence, we have
\begin{equation}
\lim_{\sigma_1\to0}
w_{\rm raw}^{\ue}
=
-\frac13
+\frac{\epsilon}{3}
-\frac{2\delta}{3}
+\Order_{\rm SR}(2).
\label{eq:UEstrictIRw}
\end{equation}
At leading slow-roll order,
the ratio tends to $-1/3$.

This zero-gradient cancellation
is shared by the comoving
and uniform-density gauges,
but the first gradient coefficients are not.
Their corresponding leading pairs are
$(-10,10/3)\sigma_1^2$
and $(-18,6)\sigma_1^2$,
respectively,
in units of $\factLW$.
The UE pair instead is
$(8/\epsilon,-8/(3\epsilon))\sigma_1^2$.
Thus equality at $k=0$
does not imply equality
at finite wavelength.

The distinction can also be seen
directly from the shift.
Using the comoving result,
\begin{equation}
Q_{\cgauge}
=
-\frac{\Psi'+\cH\Psi}
{\epsilon\cH^2},
\end{equation}
the UE shift satisfies
\begin{equation}
\frac{Q_{\ue}}{Q_{\cgauge}}
=
\frac{3\epsilon\cH^2}
{k^2+3\epsilon\cH^2}.
\label{eq:QUEQCratio}
\end{equation}
Therefore
\begin{equation}
Q_{\ue}-Q_{\cgauge}
=
-Q_{\cgauge}
\frac{k^2}
{k^2+3\epsilon\cH^2}
\simeq
-Q_{\cgauge}
\frac{\sigma_1^2}{3\epsilon}.
\label{eq:QUEQCdifference}
\end{equation}
The representatives coincide
at exactly zero gradient,
while their first gradient corrections
need not produce the same 2EMT.

\subsection{Intermediate infrared and nonuniform slow-roll expansion}

For
$\epsilon,|\delta|\ll\sigma_1^2\ll1$,
the formal lowest slow-roll terms
in Eqs.~\eqref{eq:UElw00full}
and \eqref{eq:UElwiifull} are
\begin{equation}
\widehat\tau_{00}^{\ue}
\simeq
\epsilon
(\sigma_1^2+\sigma_1^4)
\factLW,
\qquad
\widehat\tau_{ii}^{\ue}
\simeq
\epsilon
\left(
-\frac13\sigma_1^2+\sigma_1^4
\right)
\factLW.
\label{eq:UEinterIRformal}
\end{equation}
These terms are not uniformly dominant
over the entire intermediate interval.

Higher slow-roll orders contain
increasing inverse powers of $\sigma_1$.
For example,
the energy density contains
$108\epsilon^2/\sigma_1^4$.
Compared with
the nominal leading term
$\epsilon\sigma_1^2$,
its relative size is
\begin{equation}
108\frac{\epsilon}{\sigma_1^6}.
\label{eq:UEnonuniformratio}
\end{equation}
Ignoring numerical coefficients,
a representative parametric condition
for suppressing this correction is
\begin{equation}
\sigma_1^6\gg\epsilon,
\qquad
\text{or equivalently}
\qquad
\sigma_1^2\gg\epsilon^{1/3}.
\label{eq:UEdeepIRcondition}
\end{equation}
This is stronger
than the basic intermediate hierarchy.
It should be viewed
as a sufficient parametric separation,
not as a sharp boundary.

Sufficiently deep
in this gradient-dominated sector,
the leading gradient terms become
\begin{equation}
\widehat\tau_{00}^{\ue}
\simeq
\epsilon\sigma_1^2\factLW,
\qquad
\widehat\tau_{ii}^{\ue}
\simeq
-\frac13
\epsilon\sigma_1^2\factLW,
\qquad
w_{\rm raw}^{\ue}
\simeq-\frac13.
\label{eq:UEdeepinterIR}
\end{equation}
This is the regime
where the UE result approaches
the leading spatially-flat behavior.

As $\sigma_1$ decreases,
the inverse-$\sigma_1$ terms grow.
Several nominally higher slow-roll terms
then become comparable to,
or larger than,
the lowest-order contribution.
Near the denominator scale
\begin{equation}
\sigma_1^2\sim3\epsilon,
\label{eq:UEdenomscale}
\end{equation}
the wavelength and slow-roll scales
also compete directly
inside the UE constraint.
No finite slow-roll-first truncation
is uniformly reliable
through this transition region.
The full rational expressions
must be used there.

Although $w_{\rm raw}^{\ue}$
tends to $-1/3$
in both asymptotic sectors,
this does not imply
that the ratio remains $-1/3$
through the intervening transition.

\section{Uniform-expansion 2EMT in the short-wavelength domain}
\label{sec:UESW}

For $k\gg\cH$,
the short-wavelength mode
in Eq.~\eqref{eq:PsiSolutions}
is oscillatory.
We retain the slowly varying envelope
during differentiation,
and then apply
the phase average
in Eq.~\eqref{eq:phaseaverage}.
We write
\begin{equation}
C_k\equiv
|c_1|^2+|c_2|^2,
\end{equation}
and define
\begin{equation}
\factSW\equiv
\frac{2\pi Gk^4}{a^2}C_k.
\label{eq:FSWdef}
\end{equation}
The exact UE expressions,
their wavelength-first strict-UV expansions,
and their slow-roll-first intermediate-UV forms
are given below.

\begin{align}
\widehat\tau_{00}^{\mathrm{SW}}={}&
\frac{2\pi Gk^2C_k}{a^2(k^2+3\epsilon\cH^2)^4}
\Bigg\{2k^{10}+\cH^2k^8(1+31\epsilon+3\delta)
+\cH^4k^6[\delta(\delta-1)^2(\delta+3)+202\epsilon^2-2\epsilon(7\delta^2-10\delta+12)]
\nonumber\\
&\hspace{18mm}
+3\cH^6\epsilon k^4[6\delta^4-10\delta^3+56\delta^2-130\delta+78+212\epsilon^2+\epsilon(-47\delta^2+15\delta-88)]
\nonumber\\
&\hspace{18mm}
+9\cH^8\epsilon k^2[100\epsilon^3-2\epsilon^2(24\delta^2-5\delta+50)
+\epsilon(9\delta^4-27\delta^3+163\delta^2-257\delta+112)-12(\delta-1)^3(\delta+1)]
\nonumber\\
&\hspace{18mm}
+81\cH^{10}\epsilon^2[5\epsilon^3-\epsilon^2(\delta+3)^2+2\epsilon(7\delta^2-6\delta-1)-8(\delta-1)^3]
\nonumber\\
&\hspace{18mm}
-\cH^2k^2\xi_3(k^2+3\epsilon\cH^2)^2[k^2+\cH^2(\delta-1)^2]\Bigg\}\nonumber\\[1mm]
\simeq{}&
\frac{2\pi G C_k\cH^4}{a^2}
\Bigg\{
2\sigma_2^{-4}+\sigma_2^{-2}\Big[1+(3\delta+7\epsilon)-\xi_3\Big]
\nonumber\\
&\hspace{13mm}
+ \Big[ 3 \delta - 36 \epsilon +\bigl(- 5 \delta^{2} - 16 \delta \epsilon + 10 \epsilon^{2} - \xi_3\bigr) +\bigl(\delta^{3} - 14 \delta^{2} \epsilon + 2 \delta \xi_3 + 6 \epsilon \xi_3\bigr) +\bigl(\delta^{4} - \delta^{2} \xi_3\bigr) \Big]\nonumber\\
&\hspace{13mm}
+ \sigma_2^{2}\Big[ 234 \epsilon +\bigl(- 426 \delta \epsilon + 114 \epsilon^{2}\bigr) +\bigl(228 \delta^{2} \epsilon + 75 \delta \epsilon^{2} - 78 \epsilon^{3} + 6 \epsilon \xi_3\bigr)
\nonumber\\
&\hspace{18mm}
+\bigl(- 42 \delta^{3} \epsilon + 27 \delta^{2} \epsilon^{2} - 12 \delta \epsilon \xi_3 - 27 \epsilon^{2} \xi_3\bigr) +\bigl(6 \delta^{4} \epsilon + 6 \delta^{2} \epsilon \xi_3\bigr) \Big]\nonumber\\
&\hspace{13mm}
+ \sigma_2^{4}\Big[ 108 \epsilon +\bigl(- 216 \delta \epsilon - 1800 \epsilon^{2}\bigr) +\bigl(2637 \delta \epsilon^{2} - 432 \epsilon^{3}\bigr)
\nonumber\\
&\hspace{18mm}
+\bigl(216 \delta^{3} \epsilon - 999 \delta^{2} \epsilon^{2} - 270 \delta \epsilon^{3} + 378 \epsilon^{4} - 27 \epsilon^{2} \xi_3\bigr)
\nonumber\\
&\hspace{18mm}
+\bigl(- 108 \delta^{4} \epsilon + 207 \delta^{3} \epsilon^{2} + 54 \delta \epsilon^{2} \xi_3 + 108 \epsilon^{3} \xi_3\bigr) +\bigl(- 45 \delta^{4} \epsilon^{2} - 27 \delta^{2} \epsilon^{2} \xi_3\bigr) \Big]\nonumber\\
&\hspace{13mm}
+ \sigma_2^{6}\Big[ - 648 \epsilon^{2} +\bigl(648 \delta \epsilon^{2} + 8802 \epsilon^{3}\bigr) +\bigl(1944 \delta^{2} \epsilon^{2} - 9936 \delta \epsilon^{3} + 2106 \epsilon^{4}\bigr)
\nonumber\\
&\hspace{18mm}
+\bigl(- 3240 \delta^{3} \epsilon^{2} + 1350 \delta^{2} \epsilon^{3} + 189 \delta \epsilon^{4} - 1566 \epsilon^{5} + 108 \epsilon^{3} \xi_3\bigr)
\nonumber\\
&\hspace{18mm}
+\bigl(1296 \delta^{4} \epsilon^{2} - 324 \delta^{3} \epsilon^{3} - 27 \delta^{2} \epsilon^{4} - 216 \delta \epsilon^{3} \xi_3 - 405 \epsilon^{4} \xi_3\bigr) +\bigl(108 \delta^{4} \epsilon^{3} + 108 \delta^{2} \epsilon^{3} \xi_3\bigr) \Big]\nonumber\\
&\hspace{13mm}
+ \sigma_2^{8}\Big[ 1944 \epsilon^{3} +\bigl(3888 \delta \epsilon^{3} - 33696 \epsilon^{4}\bigr) +\bigl(- 23328 \delta^{2} \epsilon^{3} + 22599 \delta \epsilon^{4} - 11340 \epsilon^{5}\bigr)
\nonumber\\
&\hspace{18mm}
+\bigl(27216 \delta^{3} \epsilon^{3} + 13527 \delta^{2} \epsilon^{4} + 5508 \delta \epsilon^{5} + 5994 \epsilon^{6} - 405 \epsilon^{4} \xi_3\bigr)
\nonumber\\
&\hspace{18mm}
+\bigl(- 9720 \delta^{4} \epsilon^{3} - 2835 \delta^{3} \epsilon^{4} - 1458 \delta^{2} \epsilon^{5} + 810 \delta \epsilon^{4} \xi_3 + 1458 \epsilon^{5} \xi_3\bigr) +\bigl(405 \delta^{4} \epsilon^{4} - 405 \delta^{2} \epsilon^{4} \xi_3\bigr) \Big]\nonumber\\
&\hspace{13mm}
+ \sigma_2^{10}\Big[ - 58320 \delta \epsilon^{4} + 104490 \epsilon^{5} +\bigl(174960 \delta^{2} \epsilon^{4} + 15066 \delta \epsilon^{5} + 59778 \epsilon^{6}\bigr)
\nonumber\\
&\hspace{18mm}
+\bigl(- 174960 \delta^{3} \epsilon^{4} - 145800 \delta^{2} \epsilon^{5} - 53217 \delta \epsilon^{6} - 21870 \epsilon^{7} + 1458 \epsilon^{5} \xi_3\bigr)
\nonumber\\
&\hspace{18mm}
+\bigl(58320 \delta^{4} \epsilon^{4} + 32562 \delta^{3} \epsilon^{5} + 16767 \delta^{2} \epsilon^{6} - 2916 \delta \epsilon^{5} \xi_3 - 5103 \epsilon^{6} \xi_3\bigr)
\nonumber\\
&\hspace{18mm}
+\bigl(- 6318 \delta^{4} \epsilon^{5} + 1458 \delta^{2} \epsilon^{5} \xi_3\bigr) \Big] +\Order(\sigma_2^{12}) \Bigg\} \qquad\text{(strict UV)} \nonumber\\[1mm]\simeq{}& \frac{2\pi G C_k\cH^4}{a^2} \Bigg\{ \left[ 2\sigma_2^{-4} +\sigma_2^{-2} \right]\nonumber\\
&\hspace{13mm}
+\left[
(3\delta+7\epsilon)\sigma_2^{-2}
+(3\delta-36\epsilon)
+234\epsilon\sigma_2^2
+108\epsilon\sigma_2^4
\right]
+\Order_{\rm SR}(2;\sigma_2)
\Bigg\}
\qquad\text{(intermediate UV)}.
\label{eq:UEsw00full}
\end{align}

\begin{align}
\widehat\tau_{ii}^{\mathrm{SW}}={}&
\frac{2\pi Gk^2C_k}{3a^2(k^2+3\epsilon\cH^2)^5}
\Bigg\{2k^{12}+\cH^2k^{10}(-1+29\epsilon-\delta-4\delta^2)
+3\cH^4k^8[\delta(\delta-3)(\delta-1)^2+39\epsilon^2+\epsilon(-30\delta^2-5\delta+31)]
\nonumber\\
&\hspace{18mm}
-9\cH^6\epsilon k^6[\delta^4-69\delta^3+201\delta^2-287\delta+130+16\epsilon^2+\epsilon(39\delta^2+39\delta-158)]
\nonumber\\
&\hspace{18mm}
-27\cH^8\epsilon k^4[82\epsilon^3+\epsilon^2(-19\delta^2+93\delta-254)+\epsilon(9\delta^4-137\delta^3+353\delta^2-443\delta+170)
\nonumber\\
&\hspace{18mm}
+4(\delta-1)^2(2\delta^3-7\delta^2-4\delta+7)]
-81\cH^{10}\epsilon^2k^2[69\epsilon^3+\epsilon^2(-43\delta^2+36\delta-127)
\nonumber\\
&\hspace{18mm}
+\epsilon(7\delta^4+65\delta^3-177\delta^2+147\delta-66)-4(\delta-1)^2(6\delta^3-11\delta^2+26\delta-17)]
\nonumber\\
&\hspace{18mm}
-243\cH^{12}\epsilon^3[19\epsilon^3+\epsilon^2(-11\delta^2-22\delta+1)+2\epsilon(16\delta^3-17\delta^2+6\delta-5)
-8(\delta-1)^2(3\delta^2-\delta-1)]
\nonumber\\
&\hspace{18mm}
+3\cH^2\xi_3(k^2+3\epsilon\cH^2)[216\cH^8\epsilon^2(\delta-1)(1+\epsilon-\delta)
+9\cH^6\epsilon k^2\{-8\delta(\delta-1)^2+\epsilon(\delta^2+30\delta-55)\}
\nonumber\\
&\hspace{18mm}
+3\cH^4\epsilon k^4(2\delta^2-4\delta+3\epsilon-22)+\cH^2k^6\{6\epsilon+(\delta-1)^2\}+k^8]\Bigg\}\nonumber\\[1mm]
\simeq{}&
\frac{2\pi G C_k\cH^4}{a^2}
\Bigg\{
\frac23\sigma_2^{-4}+\sigma_2^{-2}\Big[-\frac13-\frac{\delta+\epsilon}{3}-\frac{4\delta^2}{3}+\xi_3\Big]
\nonumber\\
&\hspace{13mm}
+ \Big[ - 3 \delta + 36 \epsilon +\bigl(7 \delta^{2} - 16 \epsilon^{2} + \xi_3\bigr) +\bigl(- 5 \delta^{3} - 10 \delta^{2} \epsilon - 2 \delta \xi_3 - 6 \epsilon \xi_3\bigr) +\bigl(\delta^{4} + \delta^{2} \xi_3\bigr) \Big]\nonumber\\
&\hspace{13mm}
+ \sigma_2^{2}\Big[ - 390 \epsilon +\bigl(906 \delta \epsilon - 36 \epsilon^{2}\bigr) +\bigl(- 708 \delta^{2} \epsilon - 87 \delta \epsilon^{2} + 42 \epsilon^{3} - 78 \epsilon \xi_3\bigr)
\nonumber\\
&\hspace{18mm}
+\bigl(282 \delta^{3} \epsilon + 153 \delta^{2} \epsilon^{2} + 12 \delta \epsilon \xi_3 + 27 \epsilon^{2} \xi_3\bigr) +\bigl(- 18 \delta^{4} \epsilon - 6 \delta^{2} \epsilon \xi_3\bigr) \Big]\nonumber\\
&\hspace{13mm}
+ \sigma_2^{4}\Big[ - 252 \epsilon +\bigl(648 \delta \epsilon + 4320 \epsilon^{2}\bigr) +\bigl(- 288 \delta^{2} \epsilon - 9333 \delta \epsilon^{2} - 324 \epsilon^{3}\bigr)
\nonumber\\
&\hspace{18mm}
+\bigl(- 432 \delta^{3} \epsilon + 6813 \delta^{2} \epsilon^{2} + 558 \delta \epsilon^{3} - 72 \delta \epsilon \xi_3 - 108 \epsilon^{4} + 387 \epsilon^{2} \xi_3\bigr)
\nonumber\\
&\hspace{18mm}
+\bigl(396 \delta^{4} \epsilon - 2547 \delta^{3} \epsilon^{2} - 864 \delta^{2} \epsilon^{3} + 144 \delta^{2} \epsilon \xi_3 + 234 \delta \epsilon^{2} \xi_3 - 108 \epsilon^{3} \xi_3\bigr)
\nonumber\\
&\hspace{18mm}
+\bigl(- 72 \delta^{5} \epsilon + 99 \delta^{4} \epsilon^{2} - 72 \delta^{3} \epsilon \xi_3 + 27 \delta^{2} \epsilon^{2} \xi_3\bigr) \Big]\nonumber\\
&\hspace{13mm}
+ \sigma_2^{6}\Big[ 1944 \epsilon^{2} +\bigl(- 3240 \delta \epsilon^{2} - 27918 \epsilon^{3}\bigr) +\bigl(- 4320 \delta^{2} \epsilon^{2} + 55296 \delta \epsilon^{3} + 1944 \epsilon^{4} - 216 \epsilon^{2} \xi_3\bigr)
\nonumber\\
&\hspace{18mm}
+\bigl(12312 \delta^{3} \epsilon^{2} - 35586 \delta^{2} \epsilon^{3} - 1377 \delta \epsilon^{4} + 1296 \delta \epsilon^{2} \xi_3 + 270 \epsilon^{5} - 756 \epsilon^{3} \xi_3\bigr)
\nonumber\\
&\hspace{18mm}
+\bigl(- 8424 \delta^{4} \epsilon^{2} + 12420 \delta^{3} \epsilon^{3} + 3591 \delta^{2} \epsilon^{4} - 1944 \delta^{2} \epsilon^{2} \xi_3 - 3024 \delta \epsilon^{3} \xi_3 + 405 \epsilon^{4} \xi_3\bigr)
\nonumber\\
&\hspace{18mm}
+\bigl(1728 \delta^{5} \epsilon^{2} - 324 \delta^{4} \epsilon^{3} + 864 \delta^{3} \epsilon^{2} \xi_3 - 108 \delta^{2} \epsilon^{3} \xi_3\bigr) \Big]\nonumber\\
&\hspace{13mm}
+ \sigma_2^{8}\Big[ - 7128 \epsilon^{3} +\bigl(- 9072 \delta \epsilon^{3} + 136080 \epsilon^{4}\bigr) +\bigl(93312 \delta^{2} \epsilon^{3} - 233847 \delta \epsilon^{4} - 4860 \epsilon^{5} + 2592 \epsilon^{3} \xi_3\bigr)
\nonumber\\
&\hspace{18mm}
+\bigl(- 150336 \delta^{3} \epsilon^{3} + 111699 \delta^{2} \epsilon^{4} - 4212 \delta \epsilon^{5} - 11664 \delta \epsilon^{3} \xi_3 - 648 \epsilon^{6} - 3483 \epsilon^{4} \xi_3\bigr)
\nonumber\\
&\hspace{18mm}
+\bigl(92664 \delta^{4} \epsilon^{3} - 33777 \delta^{3} \epsilon^{4} - 12150 \delta^{2} \epsilon^{5} + 15552 \delta^{2} \epsilon^{3} \xi_3 + 22518 \delta \epsilon^{4} \xi_3 - 1458 \epsilon^{5} \xi_3\bigr)
\nonumber\\
&\hspace{18mm}
+\bigl(- 19440 \delta^{5} \epsilon^{3} + 405 \delta^{4} \epsilon^{4} - 6480 \delta^{3} \epsilon^{3} \xi_3 + 405 \delta^{2} \epsilon^{4} \xi_3\bigr) \Big]\nonumber\\
&\hspace{13mm}
+ \sigma_2^{10}\Big[ 252720 \delta \epsilon^{4} - 537030 \epsilon^{5} +\bigl(- 933120 \delta^{2} \epsilon^{4} + 684774 \delta \epsilon^{5} - 8748 \epsilon^{6} - 19440 \epsilon^{4} \xi_3\bigr)
\nonumber\\
&\hspace{18mm}
+\bigl(1263600 \delta^{3} \epsilon^{4} - 27216 \delta^{2} \epsilon^{5} + 71685 \delta \epsilon^{6} + 77760 \delta \epsilon^{4} \xi_3 + 1458 \epsilon^{7} + 47142 \epsilon^{5} \xi_3\bigr)
\nonumber\\
&\hspace{18mm}
+\bigl(- 738720 \delta^{4} \epsilon^{4} - 36450 \delta^{3} \epsilon^{5} + 32805 \delta^{2} \epsilon^{6} - 97200 \delta^{2} \epsilon^{4} \xi_3 - 133164 \delta \epsilon^{5} \xi_3 + 5103 \epsilon^{6} \xi_3\bigr)
\nonumber\\
&\hspace{18mm}
+\bigl(155520 \delta^{5} \epsilon^{4} + 3402 \delta^{4} \epsilon^{5} + 38880 \delta^{3} \epsilon^{4} \xi_3
- 1458 \delta^{2} \epsilon^{5} \xi_3\bigr) \Big]
\nonumber\\
&\hspace{13mm}+\Order(\sigma_2^{12}) \Bigg\} \qquad\text{(strict UV)} \nonumber\\[1mm]\simeq{}& \frac{2\pi G C_k\cH^4}{a^2} \Bigg\{ \left[ \frac23\sigma_2^{-4} -\frac13\sigma_2^{-2} \right]\nonumber\\
&\hspace{13mm}
+\left[
-\frac{\delta+\epsilon}{3}\sigma_2^{-2}
+(-3\delta+36\epsilon)
-390\epsilon\sigma_2^2
-252\epsilon\sigma_2^4
\right]
+\Order_{\rm SR}(2;\sigma_2)
\Bigg\}
\qquad\text{(intermediate UV)}.
\label{eq:UEswiifull}
\end{align}

\subsection{Strict ultraviolet}

Taking $\sigma_2\to0$
at fixed admissible nonzero
slow-roll parameters gives
\begin{align}
\frac{\widehat\tau_{00}^{\ue}}{\factSW}
={}&
2+
\left[
1+(7\epsilon+3\delta)-\xi_3
\right]\sigma_2^2
+\Order(\sigma_2^4),
\nonumber\\
\frac{\widehat\tau_{ii}^{\ue}}{\factSW}
={}&
\frac23
-\frac13
\left[
1+(\epsilon+\delta)
+4\delta^2-3\xi_3
\right]\sigma_2^2
+\Order(\sigma_2^4).
\label{eq:UEstrictUVcorr}
\end{align}
Therefore, we have
\begin{equation}
\widehat\tau_{00}^{\ue}
\simeq2\factSW,
\qquad
\widehat\tau_{ii}^{\ue}
\simeq\frac23\factSW,
\qquad
w_{\rm raw}^{\ue}
\simeq\frac13.
\label{eq:UEstrictUV}
\end{equation}
The leading coefficients coincide
with the longitudinal
and spatially-flat results.

\subsection{Intermediate ultraviolet}

For
$\epsilon,|\delta|\ll\sigma_2^2\ll1$,
Eqs.~\eqref{eq:UEsw00full}
and \eqref{eq:UEswiifull}
give
\begin{align}
\frac{\widehat\tau_{00}^{\ue}}{\factSW}
\simeq{}&
2+\sigma_2^2
+(3\delta+7\epsilon)\sigma_2^2+(3\delta-36\epsilon)\sigma_2^4
+234\epsilon\sigma_2^6
+108\epsilon\sigma_2^8
+\Order_{\rm SR}(2;\sigma_2),
\label{eq:UEinterUV00}
\\
\frac{\widehat\tau_{ii}^{\ue}}{\factSW}
\simeq{}&
\frac23-\frac13\sigma_2^2
-\frac{\delta+\epsilon}{3}\sigma_2^2+(-3\delta+36\epsilon)\sigma_2^4
-390\epsilon\sigma_2^6
-252\epsilon\sigma_2^8
+\Order_{\rm SR}(2;\sigma_2).
\label{eq:UEinterUVii}
\end{align}
Thus the leading
slow-roll-independent structures are
\begin{equation}
\frac{\widehat\tau_{00}^{\ue}}{\factSW}
\simeq
2+\sigma_2^2+\cdots,
\qquad
\frac{\widehat\tau_{ii}^{\ue}}{\factSW}
\simeq
\frac23-\frac13\sigma_2^2+\cdots.
\end{equation}
They coincide with
the longitudinal and spatially-flat results
at this displayed order.

The UV expansion has no analogue
of the infrared nonuniformity.
Indeed,
\begin{equation}
k^2+3\epsilon\cH^2
=
k^2
\left(
1+3\epsilon\sigma_2^2
\right),
\end{equation}
and
$3\epsilon\sigma_2^2\ll1$
throughout both ultraviolet regimes.

\subsection{Perturbative validity}

In the ultraviolet,
Eq.~\eqref{eq:QUE} gives parametrically
\begin{equation}
Q_{\ue}
=
\Order\!\left(\frac{\Psi}{k}\right).
\end{equation}
Consequently,
\begin{equation}
|\psi_{\ue}|,
|\alpha_{\ue}|,
|k\beta_{\ue}|
=
\Order(|\Psi|).
\end{equation}
A necessary \textit{small-metric condition} is therefore
\begin{equation}
|\Psi|\ll1.
\label{eq:UEUVvalidity}
\end{equation}
Unlike the spatially-flat,
comoving,
and uniform-density gauges,
the UE gauge introduces
no additional wavelength-dependent
metric lower bound in the UV.
This condition ensures the smallness of the first-order
metric perturbations, but does not by itself guarantee
the validity of the higher-order perturbative expansion.
For the intermediate UV,
Eq.~\eqref{eq:UEUVvalidity}
must also be accompanied by
\begin{equation}
\sqrt{\epsilon},
\sqrt{|\delta|}
\ll\sigma_2\ll1.
\end{equation}

\section{Five-gauge comparison and the uniform-expansion scale transition}
\label{sec:comparison}

The comparison with the longitudinal,
spatially-flat,
comoving,
and uniform-density gauges
uses the results
of Ref.~\cite{Cho:2026gjy}.
Those results employ
the same background prescription
used in the present UE calculation.
The long-wavelength results
are summarized first in Tab.~\ref*{tab:IRfive}.

\begin{table}[!ht]
\scriptsize
\setlength{\tabcolsep}{3pt}
\begin{ruledtabular}
\begin{tabular}{c|ccc|ccc}
&\multicolumn{3}{c|}{Strict IR}
&\multicolumn{3}{c}{Intermediate IR}\\
Gauge
& $\widehat\tau_{00}/\factLW$
& $\widehat\tau_{ii}/\factLW$
& $w_{\rm raw}$
& $\widehat\tau_{00}/\factLW$
& $\widehat\tau_{ii}/\factLW$
& $w_{\rm raw}$\\
\hline
L
& $-(9\epsilon^2-3\epsilon\delta)$
& $9\epsilon^2-3\epsilon\delta$
& $-1$
& $\epsilon\sigma_1^2$
& $-\epsilon\sigma_1^2/3$
& $-1/3$\\
SF
& $-(9\epsilon^2-3\epsilon\delta)$
& $9\epsilon^2-3\epsilon\delta$
& $-1$
& $\epsilon\sigma_1^2$
& $-\epsilon\sigma_1^2/3$
& $-1/3$\\
C
& $-10\sigma_1^2$
& $10\sigma_1^2/3$
& $-1/3$
& $-10\sigma_1^2-\sigma_1^4+\cdots$
& $10\sigma_1^2/3-3\sigma_1^4+\cdots$
& $-1/3$\\
UD
& $-18\sigma_1^2$
& $6\sigma_1^2$
& $-1/3$
& $-18\sigma_1^2-31\sigma_1^4/3+\cdots$
& $6\sigma_1^2-67\sigma_1^4/9+\cdots$
& $-1/3$\\
UE
& $8\sigma_1^2/\epsilon$
& $-8\sigma_1^2/(3\epsilon)$
& $-1/3$
& $\epsilon\sigma_1^2$
& $-\epsilon\sigma_1^2/3$
& $-1/3$\\
\end{tabular}
\end{ruledtabular}
\caption{
Leading gauge-fixed scalar 2EMTs
in the strict and intermediate infrared regimes.
The strict-IR C, UD, and UE entries
retain their first nonzero gradient terms.
Their displayed $w_{\rm raw}$ values
denote the corresponding ratio limits.
The UE intermediate entry applies only
sufficiently deep in the gradient-dominated sector,
where the inverse-$\sigma_1$ corrections
are parametrically subleading.
}
\label{tab:IRfive}
\end{table}

The longitudinal and spatially-flat gauges
retain slow-roll-suppressed constant terms.
The comoving,
uniform-density,
and uniform-expansion sources instead
vanish at exactly zero gradient ($\sigma_1=0$).
Their first finite-gradient coefficients,
however,
are different.
In particular,
the UE coefficient pair
is
\begin{equation}
\left(
\frac{8}{\epsilon},
-\frac{8}{3\epsilon}
\right)
\sigma_1^2\factLW,
\end{equation}
whereas the C and UD pairs
are respectively
\begin{equation}
\left(-10,\frac{10}{3}\right)
\sigma_1^2\factLW,
\qquad
(-18,6)
\sigma_1^2\factLW.
\end{equation}
Thus the common zero-gradient cancellation
does not imply equality
of the first finite-gradient source.
The component-ratio limit
$w_{\rm raw}\to-1/3$
is also common to C, UD, and UE.
This agreement should not be confused
with agreement of the 2EMT,
since both the signs
and the slow-roll scaling
of the coefficients differ.

In the intermediate infrared,
the leading terms of the UE slow-roll-first expansion
do not dominate throughout the interval
$\epsilon,|\delta|\ll\sigma_1^2\ll1$.
Only sufficiently deep
in the gradient-dominated sector
does its leading contribution
reduce to the spatially-flat result.
Closer to the transition,
inverse powers of $\sigma_1$
promote nominally higher slow-roll terms.
The complete rational UE expression
must then be used.
Thus, the intermediate UE result in Tab.~\ref*{tab:IRfive}
is applicable only sufficiently deep in the gradient-dominated regime,
not throughout the entire interval.

The characteristic UE denominator
$k^2+3\epsilon\cH^2$
defines a structural transition scale,
\begin{equation}
k^2\sim3\epsilon\cH^2,
\qquad
\sigma_1^2\sim3\epsilon.
\label{eq:UEstructuralscale}
\end{equation}
At this scale, the gradient and background terms
in the UE constraint become comparable.
However, this structural transition should be distinguished
from the loss of accuracy of a finite slow-roll-first expansion
in the intermediate IR.
For example, comparing a leading contribution of order
$\epsilon\sigma_1^2$ with a correction of order
$\epsilon^2/\sigma_1^4$ shows that the latter remains subleading
only when $\sigma_1^6\gg\epsilon$, up to numerical coefficients.
Thus, the truncated expansion may become inaccurate
at larger values of $\sigma_1$, well before the structural
transition at $\sigma_1^2\sim3\epsilon$ is reached.

The short-wavelength comparison is in Tab.~\ref*{tab:UVfive}.
\begin{table}[!ht]
\scriptsize
\setlength{\tabcolsep}{3pt}
\begin{ruledtabular}
\begin{tabular}{c|ccc|ccc}
&\multicolumn{3}{c|}{Strict UV}
&\multicolumn{3}{c}{Intermediate UV}\\
Gauge
& $\widehat\tau_{00}/\factSW$
& $\widehat\tau_{ii}/\factSW$
& $w_{\rm raw}$
& $\widehat\tau_{00}/\factSW$
& $\widehat\tau_{ii}/\factSW$
& $w_{\rm raw}$\\
\hline
L
& $2$
& $2/3$
& $1/3$
& $2+\sigma_2^2+\cdots$
& $2/3-\sigma_2^2/3+\cdots$
& $1/3$\\
SF
& $2$
& $2/3$
& $1/3$
& $2+\sigma_2^2+\cdots$
& $2/3-\sigma_2^2/3+\cdots$
& $1/3$\\
C
& $1/\epsilon$
& $1/(3\epsilon)$
& $1/3$
& $\epsilon^{-1}(1-10\sigma_2^2+\cdots)$
& $(3\epsilon)^{-1}(1+10\sigma_2^2+\cdots)$
& $1/3$\\
UD
& $1/(3\epsilon\sigma_2^2)$
& $1/(3\epsilon\sigma_2^2)$
& $1$
& $(3\epsilon)^{-1}(\sigma_2^{-2}-13+\cdots)$
& $(3\epsilon)^{-1}(\sigma_2^{-2}-13/3+\cdots)$
& $1$\\
UE
& $2$
& $2/3$
& $1/3$
& $2+\sigma_2^2+\cdots$
& $2/3-\sigma_2^2/3+\cdots$
& $1/3$\\
\end{tabular}
\end{ruledtabular}
\caption{
Leading gauge-fixed scalar 2EMTs
in the strict and intermediate ultraviolet regimes.
The intermediate columns display
the first slow-roll-independent
finite-wavelength structures.
The UE, longitudinal,
and spatially-flat gauges
share the same displayed terms.
}
\label{tab:UVfive}
\end{table}

\indent The UE ultraviolet behavior
is therefore qualitatively simple.
Its leading strict-UV coefficients
coincide with the
longitudinal and spatially-flat values.
The same agreement persists
for the displayed
slow-roll-independent intermediate corrections.
There is no $1/\epsilon$
matter-clock enhancement,
and no additional
$1/\sigma_2^2$ density-gradient enhancement.

For completeness,
the necessary UV small-metric conditions
for all five gauges are summarized in Tab.~\ref*{tab:UVvalidityFive}.
\begin{table}[!ht]
\scriptsize
\setlength{\tabcolsep}{5pt}
\begin{ruledtabular}
\begin{tabular}{lll}
Gauge
& UV small-metric condition
& Allowed range of $\sigma_2$\\
\hline
L
& $|\Psi|\ll1$
& $0<\sigma_2\ll1$\\
SF
& $|\Psi|/\sigma_2\ll1$
& $|\Psi|\ll\sigma_2\ll1$\\
C
& $|\Psi|/(\epsilon\sigma_2^2)\ll1$
& $(|\Psi|/\epsilon)^{1/2}\ll\sigma_2\ll1$\\
UD
& $|\Psi|/(\epsilon\sigma_2^3)\ll1$
& $(|\Psi|/\epsilon)^{1/3}\ll\sigma_2\ll1$\\
UE
& $|\Psi|\ll1$
& $0<\sigma_2\ll1$\\
\end{tabular}
\end{ruledtabular}
\caption{
Necessary UV small-metric conditions
at fixed oscillation amplitude $|\Psi|$
and fixed background parameters.
These conditions ensure the smallness
of the first-order metric perturbations,
but do not by themselves guarantee the validity
of the higher-order perturbative expansion.
}
\label{tab:UVvalidityFive}
\end{table}
\FloatBarrier

\indent In the intermediate UV,
in addition to the conditions in Tab.~\ref*{tab:UVvalidityFive},
each row must also satisfy
\begin{equation}
\sqrt{\epsilon},
\sqrt{|\delta|}
\ll\sigma_2\ll1.
\end{equation}
For the spatially-flat, comoving, and uniform-density gauges,
the UV small-metric conditions in Tab.~\ref*{tab:UVvalidityFive}
impose lower bounds on $\sigma_2$ at fixed nonzero $|\Psi|$
and fixed background parameters.
Thus, the strict-UV limit $\sigma_2\to0$ eventually violates
these conditions, and the corresponding coefficients
in Tab.~\ref*{tab:UVfive} should be regarded as formal asymptotic results.
These results can be applied at finite wavelengths only
where the small-metric conditions are satisfied.

The five gauges can be classified according to their slicing conditions:
\begin{equation}
\begin{array}{ll}
\text{geometrical gauges:}
& \lgauge,\ \sfgauge,\ \ue,
\\
\text{matter-clock gauges:}
& \cgauge,\ \ud.
\end{array}
\end{equation}
The longitudinal (L), spatially-flat (SF), and uniform-expansion (UE)
gauges are defined by vanishing scalar shear, spatial curvature perturbation,
and local expansion perturbation, respectively.
In contrast, the comoving (C) and uniform-density (UD) gauges
use the inflaton and energy density as matter clocks.

The L, SF, and UE gauges exhibit the same leading 2EMT
coefficients in the gradient-dominated intermediate IR and strict UV regimes,
as well as the same displayed finite-wavelength corrections in the intermediate UV.
This agreement may be related to their common geometrical character.
However, their strict-IR behaviors differ.
The L and SF sources retain slow-roll-suppressed zero-gradient terms,
whereas the UE source vanishes at zero gradient
and its first nonzero coefficients are enhanced by $1/\epsilon$.

Unlike L and SF, UE uses the local expansion scalar
as a geometrical clock, which evolves slowly during inflation,
$\dot{\Theta}_0=-3\epsilon H^2$.
This may be related to the $1/\epsilon$ enhancement
of its first nonzero strict-IR coefficients,
analogous to, but distinct from, the matter-clock enhancements in C and UD.
Although $Q_{\ue}\to Q_{\cgauge}$ as $k\to0$,
their first finite-gradient 2EMT coefficients differ.

\subsection{Scale evolution of a fixed mode}
\label{subsec:efolds}

The hierarchy has
a direct dynamical interpretation
for a fixed comoving mode.
At Hubble crossing,
$k=a_*H_*$
and $\sigma_{1*}=1$.
Since
\begin{equation}
\frac{d\ln(aH)}{dN}
=1-\epsilon,
\end{equation}
the mode evolves as
\begin{equation}
\sigma_1(\Delta N)
=
\exp\left[
-\int_{N_*}^{N_*+\Delta N}
(1-\epsilon)dN
\right]
\simeq
e^{-(1-\epsilon)\Delta N}
\simeq
e^{-\Delta N},
\label{eq:sigmaN}
\end{equation}
for slowly varying $\epsilon$.

The denominator scale
$\sigma_1^2\sim3\epsilon$
is reached,
for approximately constant $\epsilon$,
after
\begin{equation}
\Delta N_{\rm den}
\simeq
\frac{1}{2(1-\epsilon)}
\ln\left(
\frac{1}{3\epsilon}
\right).
\label{eq:Ncross}
\end{equation}
For
$\epsilon=10^{-2},10^{-3},10^{-4}$,
this gives approximately
$1.8$, $2.9$, and $4.1$
e-folds,
respectively.
Therefore, the long-wavelength regime $\sigma_1\ll1$
is reached within only a few e-folds after horizon crossing.

The slow-roll-first intermediate expansion
need not remain uniform
all the way to this scale.
For example,
the correction in
Eq.~\eqref{eq:UEnonuniformratio}
becomes important parametrically
when $\sigma_1^6$
is no longer large
relative to $\epsilon$.
This criterion concerns
the reliability of that truncation;
it is not a second
physical crossover scale.

A fixed mode therefore evolves
schematically as
\begin{equation}
\boxed{
\begin{gathered}
\text{Hubble exit / gradient dominated}
\\
\longrightarrow
\text{deep intermediate UE, SF-like at leading order}
\\
\longrightarrow
\text{nonuniform transition region}
\\
\longrightarrow
\text{strict-IR UE finite-gradient sector}.
\end{gathered}}
\end{equation}
The final strict-IR UE sector
has the enhanced coefficient
$\sigma_1^2/\epsilon$,
while still vanishing
as $\sigma_1\to0$.

\section{Interpretation and observability}
\label{sec:discussion}

The UE calculation
has three structural features.
First,
the temporal gauge condition
is differential
for $Q$ in configuration space,
but algebraic
for each Fourier mode.
Second,
the denominator
$k^2+3\epsilon\cH^2$
makes the wavelength-first
and slow-roll-first expansions
genuinely noncommuting.
Third, in the strict infrared,
the zero-gradient UE source vanishes,
while its first finite-gradient term
is enhanced by $1/\epsilon$.

The zero-gradient behavior
requires a careful interpretation.
The UE shift $Q_{\ue}$ tends
to the comoving shift $Q_{\cgauge}$
as $k\to0$,
but this equality
does not fix
the first gradient coefficient
of the quadratic source.
The C, UD, and UE sources
all vanish at exactly zero gradient,
yet their leading
$\sigma_1^2$ coefficients
are different.
The strict-IR ratio
$w_{\rm raw}\to-1/3$
therefore carries much less information
than the individual components.

The intermediate infrared
also illustrates
why asymptotic ordering matters.
At formal lowest slow-roll order,
the UE source has
the same leading gradient structure
as the spatially-flat source.
This correspondence is reliable
only sufficiently deep
in the gradient-dominated sector.
Toward the transition,
higher slow-roll terms
with inverse powers of $\sigma_1$
become comparable to,
or larger than,
the nominal leading term.
The full rational result,
rather than a finite reordered series,
then controls the UE source.

In the ultraviolet,
the UE behavior is more direct.
The UE metric perturbations
remain of order $|\Psi|$,
so the necessary small-metric condition
is simply $|\Psi|\ll1$.
The leading UE 2EMT
coincides with the
longitudinal and spatially-flat result,
without the matter-clock
$1/\epsilon$ enhancement
of the comoving gauge
or the additional
density-gradient enhancement
of the uniform-density gauge.

These results also clarify
the status of $Q$.
In the unreduced derivative form,
$Q$ remains because
the UE condition is differential
in configuration space.
In Fourier space,
Eq.~\eqref{eq:QUE}
eliminates $Q$
for each mode.
The resulting expression
may then be written in terms
of the gauge-invariant potential $\Psi$.
This is a useful representation,
but it does not by itself
assign an additional
gauge-invariance property
to the gauge-fixed 2EMT.
Equivalent gauge-fixed variables
describe the same selected representation.

The gauge-fixed 2EMT
should also not be confused
with a fully specified observable.
A physical measurement requires
an observer congruence,
a physical hypersurface,
and a measurement
or coarse-graining prescription.
An observer-centered construction
could involve quantities such as
$\tau_{\mu\nu}^{(2)}u^\mu u^\nu$
together with clock
and observer corrections.
A unique relational observable
lies beyond the present work.
Here we isolate
the gauge-fixed quadratic source
and its scale-dependent structure.

\section{Conclusions}
\label{sec:conclusion}

We have calculated
the scalar quadratic effective
energy-momentum tensor
on uniform-expansion hypersurfaces
during slow-roll inflation.
The calculation first uses
the exact mutually consistent
background identities,
and only afterward applies
the slow-roll truncation.
This prescription is the same
as that used
in the four-gauge comparison
of Ref.~\cite{Cho:2026gjy}.

The complete scalar UE gauge
is defined by
$\kappa=0$ and $E=0$.
Its temporal constraint is differential
in configuration space,
whereas in Fourier space
\begin{equation*}
Q_{\ue}(k)
=
-\frac{3(\Psi'+\cH\Psi)}
{k^2+3\epsilon\cH^2}.
\end{equation*}
The denominator
$k^2+3\epsilon\cH^2$
produces inequivalent
wavelength-first
and slow-roll-first
asymptotic expansions.

In the strict infrared,
the UE source
has no zero-gradient term.
Its first nonzero contributions are
\begin{equation}
\frac{\widehat\tau_{00}^{\ue}}{\factLW}
\simeq
\frac{8}{\epsilon}\sigma_1^2,
\qquad
\frac{\widehat\tau_{ii}^{\ue}}{\factLW}
\simeq
-\frac{8}{3\epsilon}\sigma_1^2,
\label{eq:UEconclusionIR}
\end{equation}
at leading slow-roll order.
Thus
$w_{\rm raw}^{\ue}\to-1/3$
as $\sigma_1\to0$.
The comoving and uniform-density sources
also vanish at zero gradient,
but their first gradient coefficients
are different.
Consequently,
the equality
$Q_{\ue}\to Q_{\cgauge}$
at zero gradient
does not imply equality
of the corresponding
finite-gradient 2EMTs.

In the intermediate infrared,
the formal lowest slow-roll contribution
is
\begin{equation}
\widehat\tau_{00}^{\ue}
\simeq
\epsilon
(\sigma_1^2+\sigma_1^4)
\factLW,
\qquad
\widehat\tau_{ii}^{\ue}
\simeq
\epsilon
\left(
-\frac13\sigma_1^2+\sigma_1^4
\right)
\factLW.
\end{equation}
Only sufficiently deep
in the gradient-dominated sector
do the leading terms reduce to
the spatially-flat form
with $w_{\rm raw}^{\ue}\simeq-1/3$.
As the transition is approached,
inverse powers of $\sigma_1$
promote higher slow-roll orders.
No finite slow-roll-first truncation
is uniformly reliable there,
and the full rational expressions
must be used.

In the strict ultraviolet,
the UE source becomes
\begin{equation}
\widehat\tau_{00}^{\ue}
\simeq
2\factSW,
\qquad
\widehat\tau_{ii}^{\ue}
\simeq
\frac23\factSW,
\qquad
w_{\rm raw}^{\ue}
\simeq
\frac13.
\end{equation}
The leading intermediate-UV
finite-wavelength structure
also agrees
with the longitudinal
and spatially-flat gauges.
The UE reconstructed metric
requires only
the necessary condition
$|\Psi|\ll1$,
with no additional
wavelength-dependent UV lower bound.

The five-gauge comparison reveals a regime-dependent pattern
associated with the choice of slicing.
The geometrical gauges L, SF, and UE share the leading 2EMT coefficients
in the gradient-dominated intermediate IR and strict UV,
as well as the displayed finite-wavelength corrections in the intermediate UV.
However, their strict-IR behaviors differ.
The $1/\epsilon$ enhancement of the first nonzero UE coefficients
may be related to the slow evolution of its geometrical expansion clock,
analogous to, but distinct from, the matter-clock effects in C and UD.
Although $Q_{\ue}\to Q_{\cgauge}$ as $k\to0$,
their first finite-gradient 2EMT coefficients differ.
The UE constraint also defines a structural transition scale
$k^2\sim3\epsilon\cH^2$, distinct from the scale
at which a truncated intermediate-IR expansion may lose accuracy.

We examine the gauge dependence of the general scalar quadratic 2EMT
before imposing any gauge condition.
Under an arbitrary scalar gauge transformation, its $00$ and diagonal
spatial components acquire nonvanishing variations.
In the Fourier representation, the spatial integration of quadratic products
pairs each mode $\mathbf{k}$ with $-\mathbf{k}$,
which we refer to as Fourier-mode pairing.
Under symmetric momentum integration, terms odd in $\mathbf{k}$ vanish,
whereas even terms generally survive.
The calculated gauge variations contain nonvanishing even-in-$k$ contributions,
demonstrating that the general scalar 2EMT remains gauge dependent
after Fourier-mode pairing.
This explicit transformation test thus settles the gauge-invariance question
for the general scalar quadratic 2EMT considered here,
a long-standing issue in the cosmological backreaction literature
\cite{Mukhanov:1996ak,Abramo:1997hu,Unruh:1998ic,Ishibashi:2005sj,Green:2010qy}.

In contrast, complete UE gauge fixing removes the residual scalar
gauge freedom and selects a unique representative of each gauge orbit
under the assumed boundary conditions.
This uniqueness does not imply gauge invariance of the 2EMT.

Finally,
a gauge-fixed 2EMT
is not by itself
a fully specified observable.
A relational observable
requires additional physical information
about the hypersurface,
observer,
and measurement prescription.
The present results provide
the UE source
and its scale-dependent asymptotic structure
for such future constructions.

The gauge classification also suggests
a natural next step.
In analogy with comparisons
within the matter-clock sector,
the UE slicing can be compared
with another geometrically defined clock.
A particularly natural counterpart
is uniform spacetime curvature,
$\delta R=0$,
where $R$ is the four-dimensional Ricci scalar.
Such a comparison can test
whether the scale-transition structure found here
is specific to local expansion
or is more general
among geometrical-clock gauges.

\section*{Acknowledgments}
The author is grateful to Jaichan Hwang for very helpful discussions.
The author used ChatGPT (OpenAI)
as an assistive tool during the analysis
and preparation of this manuscript.
All AI-assisted content was critically evaluated
and verified by the author,
who takes full responsibility for the
scientific content and conclusions.

\appendix
\section{General scalar-gauge variation of the 2EMT before gauge fixing}
\label{app:rawGT}

We start from the general scalar quadratic source $T_{\mu\nu}^{(2,\mathrm{eff})}$ underlying Eqs.~(A2) and (A4)
of the uniform-density companion paper Ref.~\cite{Cho:2026gjy}. We apply the arbitrary scalar transformation
$\xi^\mu=(\xi^0,\partial^i\zeta)$ directly to the perturbation variables in configuration space, substitute the
transformed variables into the general quadratic source before gauge fixing, and subtract the original expression.
No gauge condition or residual-gauge restriction is imposed. We use
$V_1\equiv V_\phi(\phi_0)$ and $V_2\equiv V_{\phi\phi}(\phi_0)$.

In simplifying the resulting expressions, contributions that are odd under reversal of the integration momentum are
omitted, since they vanish under symmetric momentum integration. In the present calculation, the total-spatial-derivative
contributions that are removed in this way belong to this odd sector. The expressions displayed below retain
nonvanishing even-in-$k$ contributions. Since an odd contribution cannot cancel a nonzero even contribution identically,
these surviving terms are sufficient to establish the nonvanishing general gauge variation discussed in
Sec.~\ref{subsec:rawGT}. We use the symbol $\doteq$ below to denote equality after omission of these $k$-odd
contributions. The spatial off-diagonal terms require an additional isotropic angular integration and are discussed
separately in Sec.~\ref{subsec:rawGT}. The resulting $00$ and diagonal spatial variations are (with no sum over $i$)
{\small
\begin{align}
\Delta_\xi\mathcal S_{00}\doteq{}&2\cH\Big[2\cH\boldsymbol\nabla\xi^0\!\cdot\!\boldsymbol\nabla E'
 +(2\cH\boldsymbol\nabla\zeta+\boldsymbol\nabla\xi^0)\!\cdot\!(\boldsymbol\nabla Q'-\boldsymbol\nabla\xi^{0\prime})\Big]
 -4\cH\boldsymbol\nabla\xi^0\!\cdot\!(\cH'\boldsymbol\nabla\zeta+\cH\boldsymbol\nabla\zeta')
\nonumber\\
&+2\cH\boldsymbol\nabla Q\!\cdot\!(2\cH'\boldsymbol\nabla\zeta+2\cH\boldsymbol\nabla\zeta'+\boldsymbol\nabla\xi^{0\prime})
 +4\cH\boldsymbol\nabla E\!\cdot\!(\cH'\boldsymbol\nabla\xi^0+\cH\boldsymbol\nabla\xi^{0\prime})
\nonumber\\
&+\nabla^2\zeta'\Big[2\cH(\nabla^2\zeta-\nabla^2E)-(\cH'+2\cH^2)\zeta'\Big]
 +\nabla^2E'\Big[-2\cH\nabla^2\zeta+(\cH'+2\cH^2)\zeta'\Big]
 +(\cH'+2\cH^2)E'\nabla^2\zeta'
\nonumber\\
&-(\cH'+2\cH^2)(\xi^{0\prime})^2
 +2\xi^0\xi^{0\prime}\Big[-\cH\cH'+4\cH^3+4\pi Ga^2V_1\phi_0'\Big]
\nonumber\\
&+(\xi^0)^2\Big[-3\cH'^{\,2}+8\cH^2\cH'+4\cH^4+a^2V_2(\cH^2-\cH')
 +4\pi Ga^4V_1^2+32\pi Ga^2V_1\cH\phi_0'\Big]
\nonumber\\
&+2\Psi\Bigg[-\nabla^4\zeta-2\cH\nabla^2\zeta'+2\cH\nabla^2\xi^0+\nabla^2\xi^{0\prime}
 +\frac{a^2V_1}{\phi_0'}\nabla^2\xi^0
 +\xi^{0\prime}\Big(\cH'+8\cH^2-\frac{a^2V_1\cH}{\phi_0'}\Big)
\nonumber\\
&\hspace{12mm}-\xi^0\Bigg(-4\cH\cH'+4\cH^3+a^2V_2\cH
 +\frac{a^2V_1(\cH'+6\cH^2)}{\phi_0'}+\frac{a^4V_1^2\cH}{\phi_0'^{\,2}}
 +8\pi Ga^2V_1\phi_0'\Bigg)\Bigg]
\nonumber\\
&+2\Psi'\Bigg[2\cH\nabla^2\zeta-\nabla^2\zeta'+\nabla^2\xi^0
 -\xi^0\Big(3\cH'+a^2V_2+\frac{7a^2V_1\cH}{\phi_0'}+\frac{a^4V_1^2}{\phi_0'^{\,2}}\Big)
 -\frac{a^2V_1}{\phi_0'}\xi^{0\prime}\Bigg]
\nonumber\\
&+2Q'\Big[(\cH'+2\cH^2)\xi^{0\prime}-\xi^0(-\cH\cH'+4\cH^3+4\pi Ga^2V_1\phi_0')\Big]
\nonumber\\
&-2Q\Bigg[\xi^{0\prime}(-\cH\cH'+4\cH^3+4\pi Ga^2V_1\phi_0')
 +\xi^0\Big(-3\cH'^{\,2}+8\cH^2\cH'+4\cH^4+a^2V_2(\cH^2-\cH')
 +4\pi Ga^4V_1^2+32\pi Ga^2V_1\cH\phi_0'\Big)\Bigg].
\label{eq:rawGT00}
\end{align}

\begin{align}
\Delta_\xi\mathcal S_{ii}\doteq{}&-\frac23(\nabla^2\zeta')^2
 +\frac23\Big[(\nabla^2E-\nabla^2\zeta)\nabla^2\zeta''+\nabla^2\zeta\,\nabla^2E''\Big]
 +2\cH(\nabla^2\zeta''-\nabla^2E'')\zeta'
\nonumber\\
&-E'\Big[2\cH\nabla^2\zeta''+(\cH'+2\cH^2)\nabla^2\zeta'\Big]
 +\frac13\nabla^2\zeta'\Big[4\cH(\nabla^2E-\nabla^2\zeta)+3(\cH'+2\cH^2)\zeta'\Big]
\nonumber\\
&+\frac13\nabla^2E'\Big[4\cH\nabla^2\zeta+4\nabla^2\zeta'+8\cH\xi^{0\prime}
 -3(\cH'+2\cH^2)\zeta'+8(\cH'+\cH^2)\xi^0\Big]
 +(2\cH^2-\cH')(\xi^{0\prime})^2
\nonumber\\
&-\frac83\xi^{0\prime}\Bigg[\cH\nabla^2\zeta'+(\cH'+\cH^2)(\nabla^2\zeta-\nabla^2E)
 +3\xi^0\Big(\frac34\cH\cH'+\cH^3+3\pi Ga^2V_1\phi_0'\Big)\Bigg]
\nonumber\\
&-\frac43\xi^0\Bigg[2(\cH'+\cH^2)\nabla^2\zeta'+\cH(\nabla^2\zeta''-\nabla^2E'')
 +4(\cH^3+2\pi Ga^2V_1\phi_0')(\nabla^2\zeta-\nabla^2E)\Bigg]
\nonumber\\
&+(\xi^0)^2\Big[-3\cH'^{\,2}-12\cH^4-a^2V_2(\cH^2-\cH')
 +4\pi Ga^4V_1^2-32\pi Ga^2V_1\cH\phi_0'\Big]
\nonumber\\
&+\frac23\boldsymbol\nabla\xi^{0\prime}\!\cdot\!(2\cH\boldsymbol\nabla\xi^0+\boldsymbol\nabla\xi^{0\prime})
 +\frac23\boldsymbol\nabla\xi^0\!\cdot\!\boldsymbol\nabla\xi^{0\prime\prime}
 +\Big[\frac43\cH(\nabla^2E-\nabla^2\zeta)+2\cH\xi^{0\prime}-2\cH'\xi^0\Big]\xi^{0\prime\prime}
\nonumber\\
&+\Psi'\Bigg[-\frac23\nabla^2\zeta'-2\xi^{0\prime\prime}
 +\frac83\Big(2\cH+\frac{a^2V_1}{\phi_0'}\Big)\nabla^2\zeta
 +\frac{6a^2V_1}{\phi_0'}\xi^{0\prime}
 +2\xi^0\Big(a^2V_2-\cH'+\frac{a^2V_1\cH}{\phi_0'}-\frac{a^4V_1^2}{\phi_0'^{\,2}}\Big)\Bigg]
\nonumber\\
&+\Psi\Bigg[-\frac43\nabla^4\zeta+\frac23\nabla^2\zeta''+\frac43\cH\nabla^2\zeta'-2\nabla^2\xi^{0\prime}
 +\frac83\Big(\cH'+2\cH^2+\frac{a^2V_1\cH}{\phi_0'}\Big)\nabla^2\zeta
 +\frac{2a^2V_1}{\phi_0'}\nabla^2\xi^0
\nonumber\\
&\hspace{12mm}-2\xi^{0\prime}\Big(3\cH'+4\cH^2-\frac{3a^2V_1\cH}{\phi_0'}\Big)-8\cH\xi^{0\prime\prime}
 +2\xi^0\Bigg(-4\cH^3+a^2V_2\cH+\frac{a^2V_1(2\cH^2-\cH')}{\phi_0'}
 -\frac{a^4V_1^2\cH}{\phi_0'^{\,2}}-8\pi Ga^2V_1\phi_0'\Bigg)\Bigg]
\nonumber\\
&-\frac43\boldsymbol\nabla Q'\!\cdot\!(\cH\boldsymbol\nabla\xi^0+\boldsymbol\nabla\xi^{0\prime})
 -\frac23\boldsymbol\nabla\xi^0\!\cdot\!\boldsymbol\nabla Q''
 -\frac23\boldsymbol\nabla Q\!\cdot\!(2\cH\boldsymbol\nabla\xi^{0\prime}+\boldsymbol\nabla\xi^{0\prime\prime})
\nonumber\\
&+\Big(\frac43\cH\nabla^2\zeta-2\cH\xi^{0\prime}+2\cH'\xi^0\Big)Q''
\nonumber\\
&+Q'\Bigg[\frac83\cH\nabla^2\zeta'+\frac83(\cH'+\cH^2)\nabla^2\zeta
 -2(2\cH^2-\cH')\xi^{0\prime}-2\cH\xi^{0\prime\prime}
 +2\xi^0(3\cH\cH'+4\cH^3+12\pi Ga^2V_1\phi_0')\Bigg]
\nonumber\\
&+Q\Bigg[\frac83(\cH'+\cH^2)\nabla^2\zeta'
 +\frac43\cH\Big[(4\cH^2+8\pi Ga^2V_1\phi_0')\nabla^2\zeta+\nabla^2\zeta''\Big]
 +2\xi^{0\prime}(3\cH\cH'+4\cH^3+12\pi Ga^2V_1\phi_0')+2\cH'\xi^{0\prime\prime}
\nonumber\\
&\hspace{12mm}+2\xi^0\Big[3\cH'^{\,2}+12\cH^4+a^2V_2(\cH^2-\cH')
 -4\pi Ga^4V_1^2+32\pi Ga^2V_1\cH\phi_0'\Big]\Bigg].
\label{eq:rawGTii}
\end{align}
}

\section{Uniform-expansion 2EMT before eliminating $Q$}
\label{app:unreduced}

For completeness, we record the uniform-expansion expressions obtained from the general Fourier-space scalar 2EMT,
Eqs.~(A2) and (A4) of the uniform-density companion paper Ref.~\cite{Cho:2026gjy}, after imposing the two UE gauge
conditions but before eliminating $Q$ with the Fourier-space constraint.  These formulas retain the derivative
structure inside the Fourier-paired quadratic brackets.  We use $V_0\equiv V(\phi_0)$,
$V_1\equiv V_\phi(\phi_0)$, and $V_2\equiv V_{\phi\phi}(\phi_0)$.

\begin{align}
\widehat\tau_{00}^{(2)}(k)={}&
 \frac{1}{\cH^2-\cH'}
 \left\{
 \left\langle(\nabla\Psi')^2\right\rangle
 +\left\langle(\Delta\Psi)^2\right\rangle
 +\left(8\cH+\frac{2a^2V_1}{\phi_0'}\right)
 \left\langle\nabla\Psi\cdot\nabla\Psi'\right\rangle
 \right.
 \nonumber\\
&\left.
 +\left[15(\cH^2-\cH')+2\cH'+5\cH^2
 +\frac{2a^2V_1\cH}{\phi_0'}\right]
 \left\langle(\nabla\Psi)^2\right\rangle
 \right.
 \nonumber\\
&\left.
 +\left[
 (6\cH'+12\cH^2)(\cH^2-\cH')
 +\frac{4a^4G\pi V_1^2\cH^2}{\cH^2-\cH'}
 +\cH'^2+4\cH^2\cH'+a^2V_2\cH^2+4\cH^4
 \right.
 \right.
 \nonumber\\
&\left.
 \left.
 \qquad
 +\frac{2a^2V_1\cH(\cH'+2\cH^2)}{\phi_0'}
 +\frac{10a^2V_1\cH(\cH^2-\cH')}{\phi_0'}
 \right]\left\langle\Psi^2\right\rangle
 \right.
 \nonumber\\
&\left.
 +\left[
 24\cH(\cH^2-\cH')
 +\frac{8a^4G\pi V_1^2\cH}{\cH^2-\cH'}
 +6\cH\cH'+2a^2V_2\cH+12\cH^3
 \right.
 \right.
 \nonumber\\
&\left.
 \left.
 \qquad
 +\frac{2a^2V_1(\cH'+5\cH^2)}{\phi_0'}
 +\frac{10a^2V_1(\cH^2-\cH')}{\phi_0'}
 \right]\left\langle\Psi\Psi'\right\rangle
 \right.
 \nonumber\\
&\left.
 +\left[
 -9(\cH^2-\cH')
 +\frac{4a^4G\pi V_1^2}{\cH^2-\cH'}
 +a^2V_2+9\cH^2
 +\frac{6a^2V_1\cH}{\phi_0'}
 \right]\left\langle(\Psi')^2\right\rangle
 \right\}
 \nonumber\\
&+\left(4\cH'-16\cH^2+\frac{2a^2V_1\cH}{\phi_0'}\right)
 \left\langle\Psi Q'\right\rangle
 +\left(6\cH+\frac{2a^2V_1}{\phi_0'}\right)
 \left\langle\Psi'Q'\right\rangle
 \nonumber\\
&+\left[
 -26\cH\cH'
 +\frac{8a^2V_1\cH'}{\phi_0'}
 +48a^2G\pi V_0\cH+2a^2V_2\cH+8\cH^3
 +\frac{8a^4G\pi V_1^2\cH}{\cH^2-\cH'}
 \right.
 \nonumber\\
&\left.
 \qquad
 +\frac{6a^2V_1\cH^2}{\phi_0'}
 +64a^2G\pi V_1\phi_0'
 \right]\left\langle\Psi Q\right\rangle
 \nonumber\\
&+\left(
 12\cH'+2a^2V_2-6\cH^2
 +\frac{8a^4G\pi V_1^2}{\cH^2-\cH'}
 +\frac{14a^2V_1\cH}{\phi_0'}
 \right)\left\langle\Psi'Q\right\rangle
 \nonumber\\
&-\left(\cH'+2\cH^2\right)\left\langle(Q')^2\right\rangle
 +\left(-8\cH\cH'+14\cH^3+8a^2G\pi V_1\phi_0'\right)
 \left\langle QQ'\right\rangle
 \nonumber\\
&+\left(
 -3\cH'^2-a^2V_2\cH'+8\cH^2\cH'
 +a^2V_2\cH^2+4\cH^4
 +32a^2G\pi V_1\cH\phi_0'
 +4a^4G\pi V_1^2
 \right)\left\langle Q^2\right\rangle .
\end{align}

\begin{align}
\widehat\tau_{ii}^{(2)}(k)={}&
\frac{1}{\cH'-\cH^2}
\left\{
\frac{1}{3}\left\langle(\nabla\Psi')^2\right\rangle
-\left\langle(\Delta\Psi)^2\right\rangle
-\left(\frac{16}{3}\cH+\frac{2a^2V_1}{\phi_0'}\right)
\left\langle\nabla\Psi\cdot\nabla\Psi'\right\rangle
\right.
\nonumber\\
&\left.
+\left[\frac{1}{3}(5\cH'-22\cH^2)-\frac{2a^2V_1\cH}{\phi_0'}\right]
\left\langle(\nabla\Psi)^2\right\rangle
\right.
\nonumber\\
&\left.
+\left[
7(\cH')^2+4\cH^2\cH'+a^2V_2\cH^2-20\cH^4
-\frac{2a^2V_1\cH(2\cH'+\cH^2)}{\phi_0'}
+\frac{4a^4G\pi V_1^2\cH^2}{\cH'-\cH^2}
\right]
\left\langle\Psi^2\right\rangle
\right.
\nonumber\\
&\left.
+\left[
14\cH\cH'
-\frac{4a^2V_1(\cH'+2\cH^2)}{\phi_0'}
+2a^2V_2\cH-32\cH^3
+\frac{8a^4G\pi V_1^2\cH}{\cH'-\cH^2}
\right]
\left\langle\Psi\Psi'\right\rangle
\right.
\nonumber\\
&\left.
+\left[
\cH'+a^2V_2-10\cH^2
-\frac{6a^2V_1\cH}{\phi_0'}
+\frac{4a^4G\pi V_1^2}{\cH'-\cH^2}
\right]
\left\langle(\Psi')^2\right\rangle
\right\}
\nonumber\\
&+\left(8\cH'+12\cH^2-\frac{2a^2V_1\cH}{\phi_0'}\right)
\left\langle\Psi Q'\right\rangle
+\left(6\cH-\frac{2a^2V_1}{\phi_0'}\right)
\left\langle\Psi'Q'\right\rangle
\nonumber\\
&+\left[
\frac{8a^2V_1\cH'}{\phi_0'}
-2a^2V_2\cH+8\cH^3
-\frac{10a^2V_1\cH^2}{\phi_0'}
+16a^2G\pi V_1\phi_0'
-\frac{8a^4G\pi V_1^2\cH}{\cH'-\cH^2}
\right]
\left\langle\Psi Q\right\rangle
\nonumber\\
&+\left(
2\cH'-2a^2V_2
-\frac{8a^4G\pi V_1^2}{\cH'-\cH^2}
-\frac{2a^2V_1\cH}{\phi_0'}
\right)
\left\langle\Psi'Q\right\rangle
+6\cH\left\langle\Psi Q''\right\rangle
\nonumber\\
&+\cH'\left\langle(Q')^2\right\rangle
+\left(
-12\cH\cH'
+16a^2G\pi V_0\cH
-8\cH^3
-8a^2G\pi V_1\phi_0'
\right)
\left\langle QQ'\right\rangle
\nonumber\\
&+\left[
-3(\cH')^2+a^2V_2\cH'
-32a^2G\pi V_1\cH\phi_0'
-a^2V_2\cH^2-12\cH^4
+4a^4G\pi V_1^2
\right]
\left\langle Q^2\right\rangle
\nonumber\\
&-2\cH^2\left\langle QQ''\right\rangle
+2\cH\left\langle Q'Q''\right\rangle .
\end{align}

\bibliography{}

\end{document}